\documentclass[reprint,amsmath,amssymb,aps,superscriptaddress]{revtex4-1}

\usepackage{bm}
\usepackage{color}
\usepackage{comment}

\usepackage{braket}

\usepackage{graphicx}
\usepackage{dcolumn}
\usepackage{here}

\usepackage{hyperref}
\hypersetup{
     colorlinks   = true,
     linkcolor    = blue,
     citecolor    = blue,
     urlcolor     = blue
}

\usepackage{appendix}

\newcommand{\R}{\mathrm{Re}}
\newcommand{\I}{\mathrm{Im}}
\newcommand{\A}{\mathcal{A}}

\newcommand{\x}{\bm{x}}
\newcommand{\kvec}{\bm{k}}
\newcommand{\p}{\bm{p}}
\newcommand{\q}{\bm{q}}
\newcommand{\J}{\bm{J}}

\begin{document}

\preprint{APS/123-QED}

\title{Quantum theory of the Intrinsic Orbital Magnetoelectric Effect in itinerant electron systems at finite temperatures}

\author{Koki Shinada}
\email{shinada.koki.64w@st.kyoto-u.ac.jp}
\author{Akira Kofuji}
\author{Robert Peters}
\affiliation{Department of Physics, Kyoto University, Kyoto 606-8502, Japan}

\date{\today}

\begin{abstract}
Magnetization can be induced by an electric field in systems without inversion symmetry $\mathcal{P}$ and time-reversal symmetry $\mathcal{T}$. This phenomenon is called the magnetoelectric (ME) effect. The spin ME effect has been actively studied in multiferroics. The orbital ME effect also exists and has been mainly discussed in topological insulators at zero temperature. In this paper, we study the intrinsic orbital ME response in metals at finite temperature using the Kubo formula. The intrinsic response originates from the Fermi sea and does not depend on the dissipation. Especially in systems with $\mathcal{PT}$-symmetry, the extrinsic orbital ME effect becomes zero, and the intrinsic ME effect is dominant. We apply the response tensor obtained in this work to a $\mathcal{PT}$-symmetric model Hamiltonian with antiferromagnetic loop current order demonstrating that the intrinsic ME effect is enhanced around the Dirac points.
\end{abstract}

\maketitle

\section{Introduction}
The magnetoelectric (ME) effect is a phenomenon, where magnetization is generated by applying an electric field or, conversely, an electric polarization is generated by applying a magnetic field. This effect appears in various systems without inversion symmetry and time-reversal symmetry. Most research on the ME effect has focused on spin degrees of freedom, e.g., in multiferroics \cite{fiebig2005revival,wang2010multiferroic,tokura2014multiferroics,dong2015multiferroic,fiebig2016evolution}, after the first observation in an antiferromagnet, $\mathrm{Cr_2O_3}$ \cite{1571980074033746432,astrov1960magnetoelectric,PhysRevLett.6.607}. However, besides spin magnetization, orbital magnetization also contributes to the total magnetization that can be induced by an electric field. Such an orbital magnetization induced by an electric field is called the orbital ME effect.
The orbital ME effect has been mainly studied in topological insulators at zero temperature. In topological insulators, it consists of two terms, a non-topological term (Kubo term) and a topological term (the Chern-Simons term) \cite{PhysRevB.78.195424,PhysRevLett.102.146805,Malashevich_2010,PhysRevB.82.245118,PhysRevLett.112.166601}. 

The theoretical description of the orbital ME effect in metals at finite temperature is an ongoing problem because the orbital magnetic dipole moment is based on the position operator, $\hat{\bm{r}}$. Because the position operator is ill-defined in periodic systems using the Bloch basis,  the direct calculation of the orbital magnetization is complicated. This difficulty also appears when calculating the orbital magnetization in equilibrium. This problem can be solved in insulators by a formalization based on the Wannier representation  \cite{PhysRevLett.95.137205,PhysRevB.74.024408}, and by the semiclassical theory \cite{PhysRevLett.95.137204}. 
While these two approaches assume zero temperature, subsequent researchers derive equations of the orbital magnetization that are applicable even in metals at finite temperature using semiclassical theory \cite{PhysRevLett.97.026603}.
Furthermore, in a full quantum mechanical approach, in Ref.~\cite{PhysRevLett.99.197202}, the orbital magnetization $M^i$ is defined in the thermodynamic sense by a derivative of the free energy $F$ by the magnetic field $B^i$ ($M^i = \partial F/ \partial B^i$). In practice, it is calculated by the thermodynamic relation $\partial M^i /\partial \mu = \partial N/\partial B^i$. This approach avoids the use of the position operator. However, this method is unique to equilibrium and cannot be applied to nonequilibrium states, such as quantum states in the presence of a current when an electric field is applied. 

Recently, the orbital ME effect in metals at finite temperatures has been attracting attention.  It has been experimentally observed in monolayer $\mathrm{MoS_2}$ \cite{lee2017valley,PhysRevLett.123.036806} and twisted bilayer graphene \cite{doi:10.1126/science.aaw3780,he2020giant}. Theoretical research is also steadily progressing. The orbital ME effect in metals is partially derived in Refs.~\cite{yoda2015current,PhysRevLett.116.077201,PhysRevB.102.184404,PhysRevB.102.201403,PhysRevB.96.035120}. This effect is also proposed in superconductors \cite{PhysRevResearch.3.L032012,PhysRevLett.128.217703}. These researches cover the extrinsic part, which originates in the change of the Fermi distribution function by an electric field and is proportional to the relaxation time $\tau$. In general, response functions also include an intrinsic part originating in the change of the wave function by an electric field, leading to, e.g., the anomalous Hall conductivity, which is related to the Berry curvature. Thus, we can expect that there is an intrinsic orbital ME effect even in metals at finite temperature, and it is, in fact, discussed using semiclassical theory \cite{PhysRevB.103.115432,PhysRevB.103.045401}. There are still no works calculating the intrinsic orbital ME effect using a fully quantum mechanical approach. However, a fully quantum mechanical approach deriving the effect going beyond the semiclassical theory is essential to understand quantum effects and calculating the effect in interacting systems.

In this work, we derive the intrinsic orbital ME response in a fully quantum mechanical approach using the Kubo formula. In Sec.~\ref{formalization}, we discuss the formalization of the orbital ME tensor from the current-current correlation function. We derive the intrinsic orbital ME tensor and discuss the physical meaning of this formula, a relationship with the thermodynamical orbital magnetic quadrupole moment, and the symmetry constraints in Sec.~\ref{sec_IOME}. Finally, we analyze the intrinsic orbital ME response in a model with antiferromagnetic loop-current order with $\mathcal{PT}$-symmetry in Sec.~\ref{model_calculation} and conclude this paper in Sec.~\ref{conclusion}.

\section{Formalization} \label{formalization}
In this section, we will discuss the formalization of the orbital ME effect. 
To derive it, we use the fact that the orbital ME response tensor is included in the current-current correlation function $\Phi_{JJ}^{ij}(\q,\omega)$. 
In the following, we will see how to extract the orbital ME tensor from the correlation function, particularly the first-order derivative by the wave number, following Ref.~\cite{PhysRevB.82.245118}.

In linear response theory, the current-current correlation function, $\Phi^{ij}_{JJ}(\q,\omega)$, is defined using the current density $\bm{J}_{\q,\omega}$ and a monochromatic electromagnetic field $\bm{A}(\x,t) = \bm{A}_{\q,\omega} e^{-i\omega t + i \q \cdot \x}$
\begin{eqnarray}
J^i_{\q,\omega} = \Phi^{ij}_{JJ}(\q,\omega) A^j_{\q,\omega}. \label{linear}
\end{eqnarray}
Let us consider the first-order derivative by the wave number $q$ of the correlation function, $\Phi^{ij,k}_{JJ}(\omega)=\partial_{q_k}\Phi^{ij}_{JJ}(\q=0,\omega)$.
To simplify the discussion,
we split $\Phi^{ij,k}_{JJ}(\omega)$ into a time-reversal odd part $\Phi^{\mathrm{(S)}ij,k}_{JJ}(\omega)$, which is symmetric in the indices $i \leftrightarrow j$, and a time-reversal even part
$\Phi^{\mathrm{(A)}ij,k}_{JJ}(\omega)$, which is antisymmetric in the indices 
$i \leftrightarrow j$.
We will see that these two parts include the multipole response tensors, such as the ME tensor, and we will see how to decompose these parts into multipole response tensors.

At first, we consider the antisymmetric part. This part has 9 independent components. Therefore it can be decomposed by a rank-2 tensor $\alpha_{ij}$ as
\begin{subequations}
\begin{align}
&\Phi^{\mathrm{(A)}ij,k}_{JJ}(\omega)
=
i \varepsilon_{ikl} \alpha_{jl}(\omega) -i \varepsilon_{jkl} \alpha_{il}(\omega) \label{asymm_phi}, \\
& \alpha_{ij}(\omega)
=
-\frac{1}{4i} \varepsilon_{jkl} \Bigl(  
2 \Phi^{\mathrm{(A)}ik,l}_{JJ}(\omega)
-
\Phi^{\mathrm{(A)}kl,i}_{JJ}(\omega)
\Bigr).
\end{align}
\end{subequations}
Substituting Eq.~(\ref{asymm_phi}) in Eq.~(\ref{linear}) and using $\Phi^{ij}_{JJ}(\q,\omega)=\Phi^{ij,k}_{JJ}(\omega)q_k$ for small wave numbers,
Eq.~(\ref{asymm_phi}) can be written as
\begin{subequations}
\begin{align}
&\J^{\mathrm{(A)}}_{\q,\omega} = \J^{\mathrm{(A)}\bm{B}}_{\q,\omega} + i\q \times \bm{M}^{\mathrm{(A)}\bm{E}}_{\q,\omega} \\
&J^{\mathrm{(A)}\bm{B} i}_{\q,\omega} = -i\omega P^{\mathrm{(A)}\bm{B} i}_{\q,\omega} =  \alpha_{ij}(\omega) B^j_{\q,\omega} \\
&M^{\mathrm{(A)}\bm{E} i}_{\q,\omega} = \frac{1}{i \omega} \alpha_{ji}(\omega) E^j_{\q,\omega}  \label{exme_effect},
\end{align}
\end{subequations}
where $\bm{E}_{\q,\omega} = i\omega \bm{A}_{\q,\omega}$ is the electric field, and $\bm{B}_{q,\omega} = i\q \times \bm{A}_{\q,\omega}$ is the magnetic field. In Ref.~\cite{PhysRevLett.116.077201}, $\alpha_{ij}$ is calculated in the range of  nonabsorbing frequencies, $\omega \ll \Delta \epsilon_{\mathrm{gap}}$. Then, $\alpha_{ij}$ can be calculated as
\begin{eqnarray}
\alpha_{ij}(\omega) = \frac{i \omega \tau}{1- i \omega \tau} \sum_{n} \int \frac{d^3k}{(2\pi)^3} \frac{\partial f(\epsilon_{n\kvec})}{\partial k_i} m_{n\kvec j} \label{ex_metensor},
\end{eqnarray}
where $\tau$ is the relaxation time, $f(\epsilon_{n\kvec})=1/(e^{\beta(\epsilon_{n\kvec}- \mu)}+1)$ is the Fermi distribution function of the $n$-th band energy $\epsilon_{n\kvec}$, and $\bm{m}_{n\kvec}$ is the spin and orbital magnetic moment.
In the long wavelength limit $\q \to 0$, only $\J^{\bm{B}}_{0,\omega}$ contributes to the transport current induced by a magnetic field. This phenomenon is called the gyrotropic magnetic effect, and  Eq.~(\ref{exme_effect}) is interpreted as the ME effect \cite{PhysRevLett.116.077201}.
This ME effect is an extrinsic part and originates from the change of the Fermi distribution function. In the following, we will see that there is another contribution derived from the symmetric part, which is the main result of this paper.

Thus, let us discuss the symmetric part $\Phi^{\mathrm{(S)}ij,k}_{JJ}(\omega)$. This part is known to generate the intrinsic nonreciprocal directional dichroism \cite{PhysRevLett.122.227402}. In the following, we will see that this part also includes the ME tensor.
The symmetric part has 18 independent components. Therefore
it can be decomposed using a traceless rank-2 tensor $\beta_{ij}$ and a totally symmetric rank-3 tensor $\gamma_{ijk}$ \cite{PhysRevB.82.245118}. $\beta_{ij}$ has 8 independent components and $\gamma_{ijk}$ has  10 independent components. Together, they have 18 independent components in total. Thus, there is a  one-to-one correspondence to the symmetric part as
\begin{subequations}
\begin{align}
& \Phi^{(\mathrm{S})ij,k}_{JJ}(\omega)
=
i \varepsilon_{jkl} \beta_{il}(\omega) + i \varepsilon_{ikl} \beta_{jl}(\omega) + \omega \gamma_{ijk}(\omega) \label{symm_phi} \\
& \beta_{li}(\omega) = \frac{1}{3i} \varepsilon_{ijk} \Phi^{(\mathrm{S})lj,k}_{JJ}(\omega) \label{beta} \\
& \gamma_{ijk}(\omega) = \frac{1}{3\omega} \Bigl( \Phi^{(\mathrm{S})ij,k}_{JJ}(\omega)+\Phi^{(\mathrm{S})jk,i}_{JJ}(\omega)+\Phi^{(\mathrm{S})ki,j}_{JJ}(\omega) \Bigr) . \label{gamma} \nonumber \\
\end{align}
\end{subequations}
Substituting Eq.~(\ref{symm_phi}) in Eq.~(\ref{linear}), we obtain
\begin{subequations}
\begin{align}\label{symm_phi_J}
&
\J^{(\mathrm{S})}_{\q,\omega}
=
\J^{(\mathrm{S}) \bm{B} \bm{E}}_{\q,\omega} + i\q \times \bm{M}^{(\mathrm{S})\bm{E}}_{\q,\omega} \\
&
J^{(\mathrm{S}) \bm{B} \bm{E}i}_{\q,\omega}
=
-i\omega \Bigl(
P^{\mathrm{(S)}\bm{B} i}_{\q,\omega}
-
i q_j Q^{\mathrm{(S)}\bm{E} ij}_{\q,\omega} 
\Bigr)  \\
& P^{\mathrm{(S)\bm{B}}i}_{\q,\omega} = \frac{1}{i\omega} \beta_{ij}(\omega) B^j_{\q,\omega} \\
& Q^{\mathrm{(S)}\bm{E}ij}_{\q,\omega} = -\frac{1}{i\omega} \gamma_{ijk}(\omega) E^k_{\q,\omega} \\
&
M^{(\mathrm{S})\bm{E}i}_{\q,\omega} = \frac{1}{i\omega} \beta_{ji}(\omega) E^{j}_{\q,\omega}.
\end{align}
\end{subequations}
In Eq.~(\ref{symm_phi_J}), $\beta_{ij}$ corresponds to the ME tensor. Thus, the ME response tensor is given by $\alpha_{ij}$ and $\beta_{ij}$.
Furthermore, since $\gamma_{ijk}$ is totally symmetric, it can only be regarded as the response of an electric quadrupole moment and cannot be included in the ME effect. Thus, $\gamma_{ijk}$ is the pure response tensor of the electric quadrupole moment induced by an electric field $Q^{\mathrm{(S)}\bm{E} ij}_{\q,\omega}$. 

The decomposition in Eq.~(\ref{symm_phi}) can also be interpreted as the expansion of the free energy by the magnetoelectric field in the path-integral sense \cite{altland_simons_2010} as
\begin{eqnarray}
F[\tilde{\bm{A}},\bm{A}] &\simeq& \tilde{A}^i_{-\q,-\omega} \Phi^{(S)ij,k}_{JJ}(\omega) q_k A^j_{\q,\omega} \nonumber \\
&=& \tilde{E}^i_{-\q,-\omega} \frac{\beta_{ij}(\omega)}{i \omega} B^j_{\q,\omega} 
+ \tilde{B}^i_{-\q,-\omega} \frac{\beta_{ji}(\omega)}{i \omega} E^j_{\q,\omega} \nonumber \\
&& + (\partial_i \tilde{E}^j)_{-\q,-\omega} \frac{- \gamma_{ijk} (\omega)}{i\omega} E^k_{\q,\omega}.
\end{eqnarray}
Here, we use $X=A, E, B$ as the external field and $\tilde{X}$ as the auxiliary field generating the observables such as the current density.
This equation shows that the expression $\beta_{ji}/i\omega$ is identical to the definition $\partial^2 F/\partial \tilde{B}^i_{-\q,\omega} \partial E^j_{\q,\omega}$ which is exactly the ME tensor. We note that in this definition, the electric field and the magnetic field are interconvertible with each other due to the Faraday law ($\bm{\nabla} \times \bm{E} + \partial \bm{B}/ \partial t =0$). Thus, when focusing on the response of the magnetization by the electric field, all components of $\partial_i E^j$, which can be transformed into a magnetic field, must be taken into account. Of course, there are components that cannot be converted into a magnetic field. These components correspond to $\gamma_{ijk}$.

First, we comment on some merits of this definition of the ME tensor. In finite systems, the naively defined polarization, magnetization, multipoles, and their response functions depend on the origin of the coordinate system \cite{10.1093/acprof:oso/9780198567271.001.0001}. However, using the current-current correlation function and the decomposition of this correlation function into the multipole response functions, we can appropriately calculate their response functions independent of the origin, even well-defined in bulk systems. In addition, the response functions are guaranteed to be gauge-invariant.

In the above discussion, the vector potential is used to describe the electric field and the magnetic field. Of course, the scalar potential also generates an electric field. In Appendix \ref{derivation_Jn}, we show that the response tensor induced by the scalar potential also includes the information on the orbital magnetic ME tensor and results in the same equations as the following discussion.

The goal of this paper is the derivation of the
orbital ME effect originating in the $\beta_{ij}$
in a uniform and static electric field.  Thus, we need to calculate $\lim_{\omega \to 0} \beta_{ij}(\omega)/i\omega$. 

\section{Derivation of the intrinsic orbital ME tensor} \label{sec_IOME}
In this section, we will derive the formula of the intrinsic orbital ME tensor at finite temperatures in periodic systems. 
In this paper, we focus on the response under a uniform and static electric field. Thus, in the final step, we will take two limits, $\q \to 0$ and $\omega \to 0$. In general, interchanging these two limits gives different results. In this paper, we consider the uniform limit (taking $\q \to 0$ before $\omega \to 0$) to calculate the dynamical response. However, we will see that the intrinsic orbital ME response originates from interband effects. Therefore, there is no need to be cautious about the order of the two limits, unlike in the extrinsic case, because there is no singularity.

\subsection{Set-up}
The Hamiltonian used in this paper is
\begin{eqnarray}
H_0 = \frac{\p^2}{2m} + V(\x) + \frac{1}{4m^2} \biggl( \frac{\partial V(\x)}{\partial \x} \times \p  \biggr) \cdot \bm{\sigma},
\end{eqnarray}
where $m$ is the mass of an electron, $V(\x) = V(\x +\bm{a})$ is a periodic potential, and $\bm{\sigma}$ is the Pauli matrix representing the spin degrees of freedom. The third term is the spin-orbital coupling term. In this paper, we use $\hbar=1$. The velocity operator $\bm{v}_0 = i [H_0 , \x]$ is
\begin{eqnarray}
\bm{v}_0 = \frac{\p}{m} + \frac{1}{4m^2} \bm{\sigma} \times \frac{\partial V(\x)}{\partial \x}.
\end{eqnarray}
Applying a monochromatic electromagnetic field $\bm{A}(\x,t) = \bm{A}_{\q,\omega} e^{-i\omega t + i\q \cdot \x}$, we find that the momentum changes as $\p \to \p + e \bm{A}(\x,t)$, where $-e$ is the charge of an electron. We here neglect the Zeeman term because we focus on the orbital magnetization. Then, the perturbed Hamiltonian in the first order of the electromagnetic field is
\begin{eqnarray}
H_{A} = \frac{e}{2} \Bigl( \bm{v}_0 \cdot \bm{A}(\x,t) + \bm{A}(\x,t) \cdot  \bm{v}_0 \Bigr) .
\end{eqnarray}
The velocity is also changed by the electromagnetic field as
$\bm{v}_{A} = i [H_A ,\x] = e \bm{A}(\x,t)/m$,
which is the diamagnetic term. The total current operator is given by $\J(\bm{r},t) = -e \{ \bm{v}_{\mathrm{tot}} , \delta(\bm{r} - \x) \}/2$, where $\bm{r}$ is just a coordinate and not an operator unlike $\x$, and $\bm{v}_{\mathrm{tot}} = \bm{v}_0 + \bm{v}_A $. This current operator satisfies the equation of continuity $\bm{\nabla}_{\bm{r}} \cdot \J(\bm{r},t) = - i[H_0 +H_A , n(\bm{r} ) ]$ , where $n(\bm{r})=-e \delta(\bm{r} - \x)$ is the electron density.
Fourier transforming this current operator yields
\begin{eqnarray}
J_{\q,\omega} &=& \int d^3r J(\bm{r},t) e^{i\omega t -i \q \cdot \bm{r}} \nonumber \\
&=&-e \biggl( \frac{1}{2} \{ \bm{v}_0 , e^{+i \omega t -i\q \cdot \x} \} +  \frac{e\bm{A}_{\q,\omega}}{m} \biggr).
\end{eqnarray}
Using the second quantization and the Kubo linear response theory, the symmetric part of the current-current correlation function $\Phi^{(\mathrm{S})ij}_{JJ}(\q,\omega)$ is given by
\begin{widetext}
\begin{eqnarray}
\Phi^{(\mathrm{S})ij}_{JJ} (\q,\omega)
&=&
-e^2 \int_{\mathrm{BZ}} \frac{d^3k}{(2\pi)^3}\sum_{mn} \frac{f(\epsilon_{n\kvec+\q/2}) - f(\epsilon_{m\kvec-\q/2})}{\epsilon_{n\kvec+\q/2} - \epsilon_{m\kvec-\q/2} -( \omega + i\delta )} 
\R \Bigl( \bra{u_{m\kvec-\q/2}} v^i_{\kvec} \ket{u_{n\kvec+\q/2}} \bra{u_{n\kvec+\q/2}} v^j_{\kvec} \ket{u_{m\kvec-\q/2}} \Bigr) \nonumber \\
&& - e^2 \int_{\mathrm{BZ}} \frac{d^3k}{(2\pi)^3}  \sum_{n} \frac{f(\epsilon_{n\kvec})}{m} \delta_{ij}.
\end{eqnarray}
Here, we use the following notations:
\begin{equation}
H_0 \ket{\psi_{n\kvec}} = \epsilon_{n\kvec} \ket{\psi_{n\kvec}},\hspace{10pt}
\ket{u_{n\kvec}} = e^{-i \kvec \cdot \x} \ket{\psi_{n\kvec}},\hspace{10pt}
H_{\kvec} = e^{-i \kvec \cdot \x} H_0 e^{i \kvec \cdot \x},\hspace{10pt}
\bm{v}_{\kvec} = \frac{\partial H_{\kvec}}{\partial \kvec} .
\end{equation}
The Bloch wave number $\kvec$ lies in the first Brillouin zone (BZ). $\epsilon_{n\kvec}$ is the eigenenergy of the $n$-th band, $\ket{\psi_{n\kvec}}$ is a Bloch function, $\ket{u_{n\kvec}}$ is the periodic part of the Bloch function, and $\bm{v}_{\kvec}$ is the velocity operator of the Bloch basis. 
Expanding $\Phi^{\mathrm{(S)}ij}_{JJ}(\q,\omega)$ by $\q$ up to the first order and using the relationship in Eq.~(\ref{beta}) and Eq.~(\ref{gamma}), we can obtain the uniform and static orbital ME tensor $\chi^{\mathrm{(me)}}_{ij}$ and
the pure electric quadrupole conductivity $\sigma^{\mathrm{(eq)}}_{ijk}$, which represents the contribution from the electric quadrupole moment to the current density. 

\subsection{Intrinsic orbital ME tensor}
In this subsection, we show the equations of the intrinsic orbital ME tensor and the pure electric quadrupole conductivity and discuss the physical meaning of these responses.
The expressions for the intrinsic orbital ME tensor and the pure electric quadrupole conductivity are (see Appendix~\ref{derivation_JJ} for a detailed derivation)
\begin{eqnarray}
&&\chi^{\mathrm{(me)}}_{ij}
=
\lim_{\omega \to 0} \R \biggl[ \frac{\beta_{ij}(\omega)}{i\omega} \biggr]
=
-e^2\int_{\mathrm{BZ}} \frac{d^3k}{(2\pi)^3}
\sum_{n} f(\epsilon_{n\kvec}) \biggl(
\frac{1}{3} \varepsilon_{klj} \partial_{l} g^{ik}_{n\kvec} - \sum_{m(\neq n)} \frac{2}{\epsilon_{nm\kvec}}
\R \Bigl[ 
\A^i_{nm} M^j_{mn}
-\frac{1}{3} \delta_{ij} \A^k_{nm} M^k_{mn}
\Bigr]
\biggr) \label{me_response} \nonumber \\ \\
&& \sigma^{\mathrm{(eq)}}_{ijk} 
= 
\lim_{\omega \to 0 } \R \gamma_{ijk}(\omega)
=
e^2 \int_{\mathrm{BZ}} \frac{d^3k}{(2\pi)^3} \sum_{n} \biggl\{
 \frac{-f'(\epsilon_{n\kvec})}{\delta^2} (\partial_i \epsilon_{n\kvec}) (\partial_j \epsilon_{n\kvec}) (\partial_k \epsilon_{n\kvec})
+
\frac{f(\epsilon_{n\kvec})}{3} \Bigl(
\partial_{i}g^{jk}_{n\kvec} +\partial_{k}g^{ij}_{n\kvec} +\partial_{j}g^{ki}_{n\kvec}
\Bigr)
\biggr\}
\label{eq_response}, \nonumber \\
\end{eqnarray}
\end{widetext}
where $\partial_l = \partial/ \partial k_l$, $\epsilon_{nm\kvec} = \epsilon_{n\kvec} - \epsilon_{m\kvec}$ and $f'(\epsilon_{n\kvec}) = \partial f(\epsilon_{n\kvec})/\partial \epsilon_{n\kvec}$. These equations are the main results of this paper. 

In the following, we discuss the physical meaning of each term in Eq.~(\ref{me_response}) and Eq.~(\ref{eq_response}).
The first term in Eq.~(\ref{me_response}) and the second term in  Eq.~(\ref{eq_response}) include $g^{ij}_{n\kvec}$, which is the quantum metric \cite{provost1980riemannian,resta2011insulating},
\begin{eqnarray}
g^{ij}_{n\kvec} = \sum_{m (\neq n)} \R [\A^i_{nm\kvec} \A^j_{mn\kvec}],
\end{eqnarray}
where $\A^i_{nm\kvec} = i \braket{u_{n\kvec} | \partial_i u_{m\kvec}}$ is the Berry connection. 
The quantum metric is a metric 
measuring the distance between wave functions on a parameter space (e.g., the Bloch wave number). This metric is an important geometric quantity characterizing quantum states in the Brillouin zone together with the Berry curvature. The quantum metric can be interpreted as an electric quadrupole moment of a wave packet \cite{PhysRevB.99.121111,PhysRevLett.122.227402}, and it also contributes to the thermodynamical electrical quadrupole moment \cite{PhysRevB.102.235149}. 
 This metric is also included in the pure electric quadrupole conductivity in the second term of Eq.~(\ref{eq_response}). This term represents the electric quadrupole moment induced by the change of the distribution function (see also Eq.~(10) in Ref.~\cite{PhysRevB.102.235149}). On the other hand, the first term in Eq.~(\ref{me_response}) comprises the antisymmetric parts between the indices of the derivative and the index of the metric.
 This term can be rewritten using integration by parts to $\varepsilon_{klj} v^0_{n\kvec} g^{ik}_{n\kvec}$
 on the Fermi surface. 
 Because the metric behaves as $g^{ij} \sim x^ix^j$ having the same symmetry as the electric quadrupole moment, this term behaves like $x^i (\bm{x} 
 \times \bm{v}^0_{n\kvec})^j$. 
 Here, $x_i$ represents the position, and
 we use $\bm{v}^0_{n\kvec} = \partial \epsilon_{n\kvec} / \partial \kvec$, 
 which is the group velocity of the $n$-th band. Thus, this term corresponds to a magnetic quadrupole moment and also appears in the thermodynamical orbital magnetic quadrupole moment \cite{PhysRevB.98.020407,PhysRevB.98.060402}, moreover, contributing to the orbital ME effect.

The second term in the Eq.~(\ref{me_response}) includes
\begin{eqnarray}
\bm{M}_{mn} = \sum_{l(\neq n)} \frac{1}{2} (\bm{v}_{ml\kvec} + \bm{v}^0_{n\kvec} \delta_{ml} ) \times \bm{\A}_{ln\kvec},
\end{eqnarray}
where $\bm{v}_{ml\kvec} = \bra{u_{m\kvec}} \bm{v}_{\kvec} \ket{u_{l\kvec}}$ is the matrix element of the velocity operator $\bm{v}_{\kvec}$. $\bm{M}_{mn}$ has the form of $\bm{v} \times \bm{r}$, so it can be interpreted as an off-diagonal element of the orbital magnetic moment. The second term in Eq.~(\ref{me_response}) can be interpreted as the perturbation of the wave function affecting the orbital magnetization $M$ created by an electric field.
Thus, this term includes the Berry connection $\A$, which acts like a polarization conjugate with the electric field. This term also appears in the orbital magnetic quadrupole moment \cite{PhysRevB.98.020407,PhysRevB.98.060402}.

Finally, we comment on the gauge invariance of the obtained equations. The parts of the equation that depend on wave functions are in the form of the off-diagonal Berry connection, $\bm{\A}_{nm\kvec}$. The off-diagonal Berry connection is gauge-invariant; therefore, these equations are also gauge-invariant. Because this approach is based on the current-current response function, the gauge invariance should be guaranteed. In addition, we calculate these tensors using a scalar potential instead of the vector potential, and we obtain the same result as  Eq.~(\ref{me_response}) and Eq.~(\ref{eq_response}) (see Appendix \ref{derivation_Jn}). Thus, these tensors do not depend on the choice of the gauge of the electromagnetic field.

\subsection{Relation with the orbital magnetic quadrupole moment and symmetry constraints}

The magnetic quadrupole moment is believed to be the origin of the ME effect.
The ME effect needs both inversion symmetry and time-reversal symmetry breaking. These conditions are the same for the emergence of the odd-parity magnetic multipole, including the magnetic quadrupole moment.
The ME tensor $\chi^{\mathrm{me}}_{ij}$, in general, can be decomposed into three terms, the magnetic monopole moment (the trace of $\chi^{\mathrm{me}}_{ij}$), the magnetic toroidal moment (the antisymmetric part of $\chi^{\mathrm{me}}_{ij}$), and the magnetic quadrupole moment (the traceless and symmetric part of $\chi^{\mathrm{me}}_{ij}$) following their symmetry \cite{PhysRevB.76.214404,spaldin2008toroidal,PhysRevB.88.094429,PhysRevB.93.195167}. This fact implies that the ME response originates from the multipole. To solidify this statement, we can show the direct relationship between the orbital magnetic quadrupole moment and the orbital ME response, known as the St\v{r}eda formula. The formula is first presented by St\v{r}eda as the relation between the quantum Hall conductivity and the orbital magnetization in insulators at zero temperature \cite{Streda_1982}. A similar relationship is also valid for the orbital ME response, as discussed in Ref.~\cite{PhysRevB.98.020407,PhysRevB.98.060402}. Our equation satisfies this relationship,
\begin{eqnarray}
\chi^{\mathrm{(me)}}_{ij} = -e \frac{\partial Q^{\mathrm{(m)}}_{ij}}{\partial \mu},
\end{eqnarray}
in insulators at zero temperature, as we can see by comparing the orbital magnetic quadrupole moment $Q^{\mathrm{(m)}}_{ij}$ in the Refs.~\cite{PhysRevB.98.060402,PhysRevB.98.020407}. 
In addition, the connection between them is also discussed in quantum wells \cite{PhysRevResearch.2.043060}, and these facts imply the possibility of the detection of orbital magnetic quadrupole moments using the orbital ME effect.

We note that our equation does not include the trace of the orbital ME tensor, which corresponds to the monopole term. This part is known to be a boundary effect and originates in the axion coupling described by the Chern-Simons action as discussed in 3-dimensional topological insulators \cite{PhysRevB.78.195424,PhysRevLett.102.146805,Malashevich_2010,PhysRevB.82.245118}. 
This problem also occurs in an identical approach at zero temperature \cite{PhysRevB.82.245118} and the thermodynamical definition of the orbital magnetic quadrupole moment, which should include the Chern-Simons term to fulfill the St\v{r}eda formula \cite{PhysRevB.98.020407}. On the other hand, the semiclassical approach in Ref.~\cite{PhysRevB.103.115432} successfully derives the Chern-Simons term in non-Chern insulators. However, the extension of this term in general metals and insulators including  Chern insulators is an ongoing problem in both the full quantum approach and the semiclassical theory.

Next, let us discuss symmetry constraints. As mentioned above, the ME effect is nonzero in systems without inversion and time-reversal symmetry. Time-reversal symmetry breaking can be satisfied in a macroscopic sense, such as a system including dissipation. 
Thus, the extrinsic ME effect in the Eq.~(\ref{ex_metensor}) can occur in the DC-limit ($\lim_{\omega \to 0} \alpha_{ij}(\omega)/i\omega$) even if the system described by the Hamiltonian $H_0$ has time-reversal symmetry. However, the intrinsic part in Eq.~(\ref{me_response}) is zero. On the other hand, if the Hamiltonian $H_0$ satisfies $\mathcal{PT}$-symmetry, which is the product of the inversion operation $\mathcal{P}$ and the time-reversal operation $\mathcal{T}$, the orbital magnetic moment $\bm{m}_{n\kvec}$ is zero resulting in a vanishing extrinsic ME effect. In this situation, only the intrinsic orbital ME effect can exist. Thus, $\mathcal{PT}$-symmetric systems, such as antiferromagnetic order, are suitable for the experimental observation of the intrinsic orbital ME effect. In Sec.~\ref{model_calculation}, we will calculate the intrinsic orbital ME effect for a model Hamiltonian with $\mathcal{PT}$-symmetry. In particular, we look at a system with antiferromagnetic loop current order purely originating from the orbital degrees of freedom.

Finally, we comment  on two previous works \cite{PhysRevB.103.115432,PhysRevB.103.045401} studying the intrinsic orbital ME effect. They derived  equations applicable to two-dimensional systems, including metals, and three-dimensional non-Chern insulators, using the semiclassical theory. This dimensional constraint is attributed to the second Chern form. In our work, we use a fully quantum mechanical approach, and our equations are applicable to arbitrary-dimensional systems, including metals. However, our formula of the intrinsic OME tensor does not include the Chern-Simons term as mentioned above. In addition, we note that there is a difference between our equation and the equation in Ref.~\cite{PhysRevB.103.115432,PhysRevB.103.045401}, i.e., the $1/3$ factor in Eq.~(\ref{me_response}) is replaced to $1/2$ in those papers.

\section{model calculation} \label{model_calculation}
\begin{figure*}[t]
\includegraphics[width=0.4\linewidth]{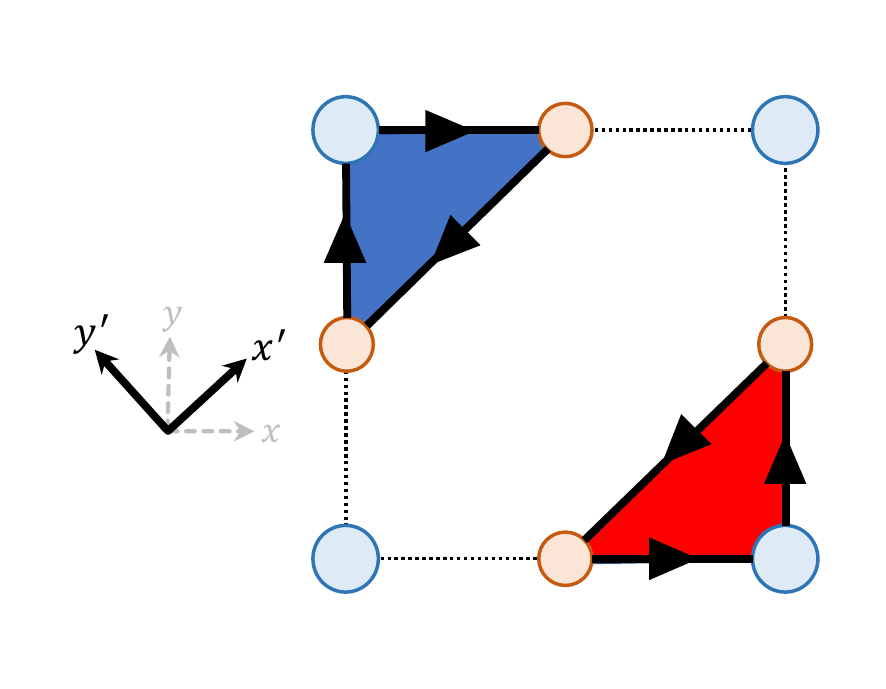}
\includegraphics[width=0.4\linewidth]{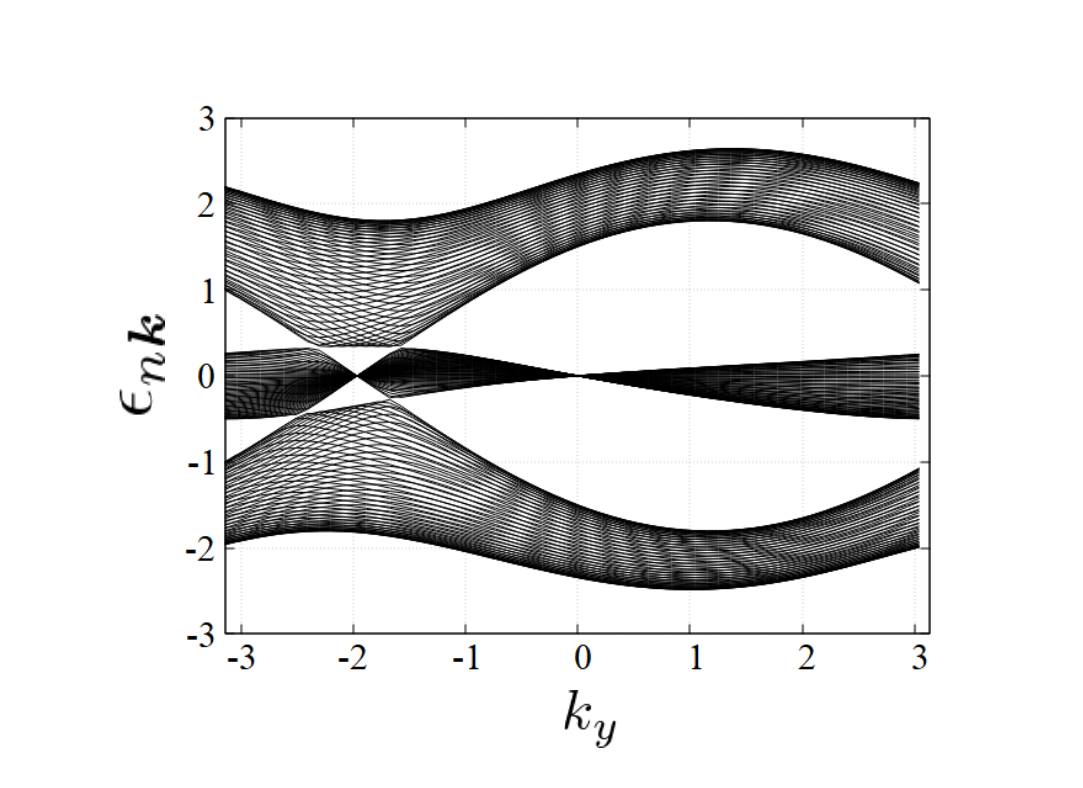} 
\caption{(Left) The loop current order on the $\mathrm{Cu}$-$\mathrm{O_2}$ plane. The blue circles are Cu atoms, and the red circles are O atoms. The blue and red shaded areas represent the local orbital magnetization with opposite signs along the $z$ axis. 
(Right) The band structure of the used model. This model has three bands and includes four Dirac points at energies $E=-0.44t$, $E=-0.27t$, and $E=0.33t$.}  \label{ybco}
\end{figure*}
\begin{figure}
\includegraphics[width=1.0\linewidth]{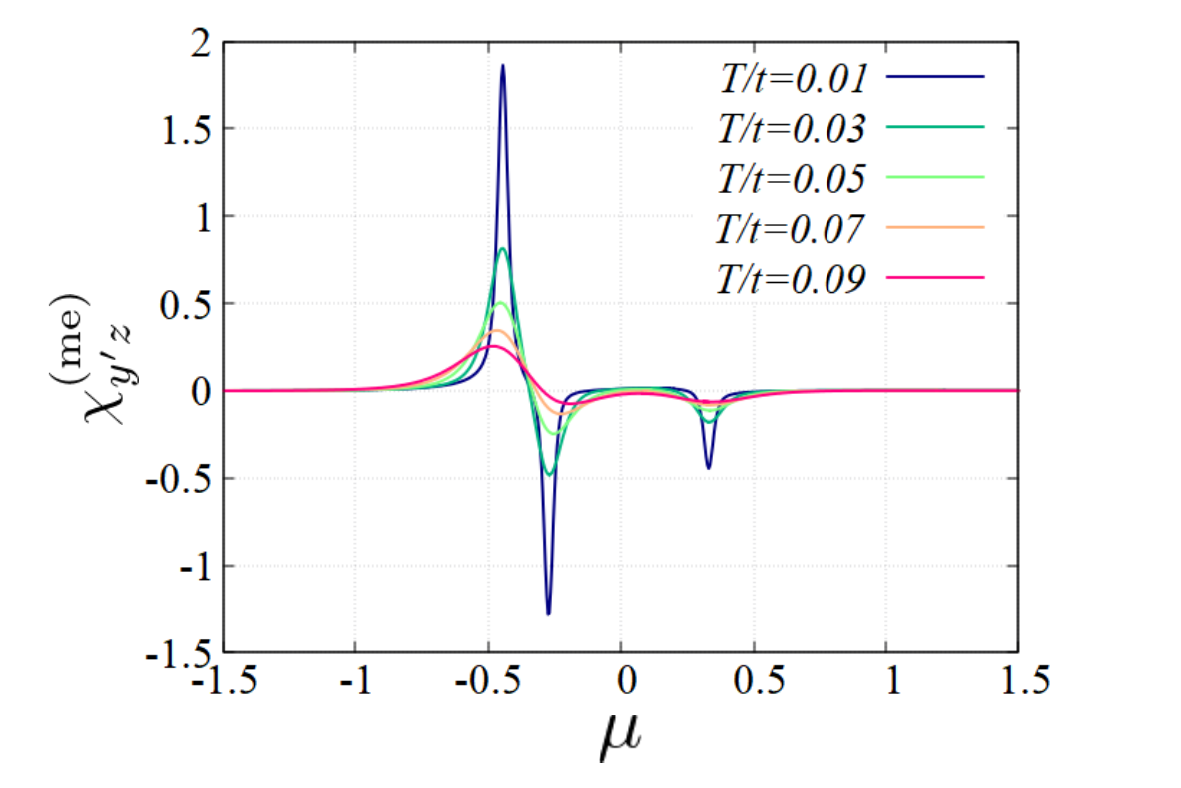}
\caption{The intrinsic orbital ME tensor $\chi^{\mathrm{(me)}}_{y'z}$ for different chemical potentials and temperatures. We set $t=1.0, r=1.5t, t'=0.5t$ for the numerical calculation, and we introduce an infinitesimal dissipation $i \delta = 0.001i$ because the response function behaves as a delta function at the Dirac points, as discussed in the main text.
We use $\mu$ in units of $t$ and $\chi^{\mathrm{(me)}}_{y'z}$ in units of $e^2 a/\hbar$.
}
\label{IOME}
\end{figure}
In this section, we calculate the intrinsic orbital ME tensor for a model Hamiltonian. Because the orbital magnetization does, in principle, not need the spin degrees of freedom, we consider here a spinless system, focusing on the contribution of the orbital moment. As discussed above, the intrinsic orbital ME effect is dominant for $\mathcal{PT}$-symmetric systems. Thus, we analyze an example with an antiferromagnetic loop current order.

Loop current order is a kind of orbital order. Intuitively, electrons rotate locally across some sites inducing local orbital magnetic moments. Loop current order has been studied in cuprates \cite{PhysRevB.55.14554,PhysRevLett.96.197001,li2008unusual,BOURGES2011461,PhysRevLett.111.047005,zhao2017global} and recently reported in the Kagome superconductors $\mathrm{AV_3Sb_5}$ ($\mathrm{A} = \mathrm{K}, \mathrm{Rb}, \mathrm{Cs}$) \cite{mielke2022time}, a Mott insulator $\mathrm{Sr_2IrO_4}$ \cite{zhao2016evidence,jeong2017time,PhysRevX.11.011021} as candidates and it is also discussed in an orbital order of the twisted bilayer graphenes \cite{liu2021orbital,PhysRevX.9.031021,PhysRevB.103.035427,PhysRevX.10.031034}.

In this section, we use a model Hamiltonian for antiferromegnatic loop current order in cuprates shown in Fig.~\ref{ybco}.
This loop current order belongs to an orbital magnetic quadrupole order.
This Hamiltonian can be written as \cite{PhysRevB.98.060402,BOURGES2011461,PhysRevB.85.155106}
\begin{eqnarray}
H_{\kvec} = 
\begin{pmatrix}
0 & its_x + ir c_x & its_y + irc_y \\
-its_x -irc_x & 0 & t' s_x s_y \\
-its_y -i rc_y & t' s_x s_y & 0 
\end{pmatrix} ,
\end{eqnarray}
where $s_i = \sin (k_i/2)$ and $c_i = \cos(k_i/2)$, and we set the lattice constant $a=1$. This Hamiltonian includes three orbitals without spin degrees of freedom. The basis is $\ket{d}$ on the copper sites, and $\ket{p_x}$ and $\ket{p_y}$ on the oxygen sites. $t$ and $t'$ are the hopping parameter and $r$ is the order parameter of the loop current in Fig.\ref{ybco}.
The band structure of this model is shown in Fig.~\ref{ybco}. This model has four Dirac points at energies $E=-0.44t$, $E=-0.27t$, and $E=0.33t$.
We rename the coordinates $(k_x,k_y)$ to $(k_{x'},k_{y'})$ for reasons of simplicity in the following discussion. 

Let us discuss the intrinsic orbital ME tensor in this model. In 2-dimensional systems, the diagonal components of $\chi^{\mathrm{(me)}}_{ij}$ vanish. Because this Hamiltonian has $y'$-mirror symmetry, the only nonzero component is $\chi^{\mathrm{(me)}}_{y'z}$. We show the result of $\chi^{\mathrm{(me)}}_{y'z}$ in Fig.~\ref{IOME} and plot the dependence on the temperature and the chemical potential. 
We here use the following parameters $t=1.0$, $t'=0.5t$, and $r=1.5t$ for the numerical calculations, and we introduce an infinitesimal dissipation $i\delta$ as an adiabatic factor.
In the result of the orbital ME effect, we can see peak structures at the energies of the Dirac points at low temperatures. This behavior is typical for a system with linear dispersion. 
When we analyze $\tilde{Q}^{\mathrm{(m)}}_{y'z}(\mu) = \int ^{\mu}_{-\infty} d\mu' \chi^{\mathrm{(me)}}_{y'z}(\mu')$, which is a part of the orbital magnetic quadrupole moment, in a Dirac Hamiltonian having the same symmetry as our model above \cite{PhysRevB.98.060402},
\begin{eqnarray}
H^{\mathrm{Dirac}}_{\kvec} = v' k_{x'} + v_x k_{x'} \sigma_{x} + v_y k_{y'} \sigma_{y},
\end{eqnarray}
 we can see that $\tilde{Q}^{\mathrm{(m)}}_{y'z}(\mu)$ shows a step function behavior and jumps from $e^2v' |v_y|/16\pi |v_x|$ to $-e^2v' |v_y|/16\pi |v_x|$ at the Dirac points with an additional logarithmic dependence (see Appendix \ref{dirac_hamiltonian}). Thus,  $\chi^{\mathrm{(me)}}_{y'z}(\mu)$ behaves at the Dirac point as a delta function.

\section{conclusion} \label{conclusion}
In summary, we have derived the intrinsic orbital ME effect within a fully quantum mechanical approach using the Kubo formula. The obtained formula is based on the current-current correlation function, so the response tensor is gauge invariant and does not depend on the origin of the coordinate system. It is well-defined as an observable quantity in bulk systems. The formula is applicable to insulators and metals at zero and finite temperatures and is valid in arbitrary dimensions.
We have shown that the intrinsic ME tensor satisfies the St\v{r}eda formula, which is the direct relationship to the thermodynamic orbital magnetic quadrupole moment in insulators at zero temperature. The intrinsic part of the orbital ME tensor is dominant in $\mathcal{PT}$-symmetric systems because the extrinsic part is zero. Thus, we have applied the obtained formula to an antiferromagnetic loop current order proposed in cuprates. We have demonstrated that the intrinsic ME tensor is strongly enhanced at low temperatures, especially around the Dirac points.

The obtained formula can also apply to other $\mathcal{PT}$-systems such as antiferromagnets. Because this formula describes the orbital magnetization induced by an electric field, it will be helpful in the study of orbital orders, e.g., loop current orders beyond spin orders. Furthermore, this formula can be used in first-principle calculations and allows for detailed calculations in real materials.

Finally, shortly before finishing our work, we noticed a related work studying the spatially dispersive natural optical conductivity \cite{pozo2022multipole}. In our paper, we focus on the orbital ME effect.

\section*{acknowledgements}
K.S. acknowledges support as a JSPS research fellow and is supported by JSPS KAKENHI, Grant No.22J23393. A.K. acknowledges support by JST, the establishment of university fellowships towards the creation of science,
technology, and innovation, Grant Number JPMJFS2123. R.P. is supported by JSPS KAKENHI No.~JP18K03511. 

\bibliography{intrinsicOME_arxiv}

\begin{thebibliography}{62}%
\makeatletter
\providecommand \@ifxundefined [1]{%
 \@ifx{#1\undefined}
}%
\providecommand \@ifnum [1]{%
 \ifnum #1\expandafter \@firstoftwo
 \else \expandafter \@secondoftwo
 \fi
}%
\providecommand \@ifx [1]{%
 \ifx #1\expandafter \@firstoftwo
 \else \expandafter \@secondoftwo
 \fi
}%
\providecommand \natexlab [1]{#1}%
\providecommand \enquote  [1]{``#1''}%
\providecommand \bibnamefont  [1]{#1}%
\providecommand \bibfnamefont [1]{#1}%
\providecommand \citenamefont [1]{#1}%
\providecommand \href@noop [0]{\@secondoftwo}%
\providecommand \href [0]{\begingroup \@sanitize@url \@href}%
\providecommand \@href[1]{\@@startlink{#1}\@@href}%
\providecommand \@@href[1]{\endgroup#1\@@endlink}%
\providecommand \@sanitize@url [0]{\catcode `\\12\catcode `\$12\catcode
  `\&12\catcode `\#12\catcode `\^12\catcode `\_12\catcode `\%12\relax}%
\providecommand \@@startlink[1]{}%
\providecommand \@@endlink[0]{}%
\providecommand \url  [0]{\begingroup\@sanitize@url \@url }%
\providecommand \@url [1]{\endgroup\@href {#1}{\urlprefix }}%
\providecommand \urlprefix  [0]{URL }%
\providecommand \Eprint [0]{\href }%
\providecommand \doibase [0]{http://dx.doi.org/}%
\providecommand \selectlanguage [0]{\@gobble}%
\providecommand \bibinfo  [0]{\@secondoftwo}%
\providecommand \bibfield  [0]{\@secondoftwo}%
\providecommand \translation [1]{[#1]}%
\providecommand \BibitemOpen [0]{}%
\providecommand \bibitemStop [0]{}%
\providecommand \bibitemNoStop [0]{.\EOS\space}%
\providecommand \EOS [0]{\spacefactor3000\relax}%
\providecommand \BibitemShut  [1]{\csname bibitem#1\endcsname}%
\let\auto@bib@innerbib\@empty
\bibitem [{\citenamefont {Fiebig}(2005)}]{fiebig2005revival}%
  \BibitemOpen
  \bibfield  {author} {\bibinfo {author} {\bibfnamefont {M.}~\bibnamefont
  {Fiebig}},\ }\href
  {https://iopscience.iop.org/article/10.1088/0022-3727/38/8/R01} {\bibfield
  {journal} {\bibinfo  {journal} {Journal of physics D: applied physics}\
  }\textbf {\bibinfo {volume} {38}},\ \bibinfo {pages} {R123} (\bibinfo {year}
  {2005})}\BibitemShut {NoStop}%
\bibitem [{\citenamefont {Wang}\ \emph {et~al.}(2010)\citenamefont {Wang},
  \citenamefont {Hu}, \citenamefont {Lin},\ and\ \citenamefont
  {Nan}}]{wang2010multiferroic}%
  \BibitemOpen
  \bibfield  {author} {\bibinfo {author} {\bibfnamefont {Y.}~\bibnamefont
  {Wang}}, \bibinfo {author} {\bibfnamefont {J.}~\bibnamefont {Hu}}, \bibinfo
  {author} {\bibfnamefont {Y.}~\bibnamefont {Lin}}, \ and\ \bibinfo {author}
  {\bibfnamefont {C.-W.}\ \bibnamefont {Nan}},\ }\href
  {https://www.nature.com/articles/am201054} {\bibfield  {journal} {\bibinfo
  {journal} {NPG asia materials}\ }\textbf {\bibinfo {volume} {2}},\ \bibinfo
  {pages} {61} (\bibinfo {year} {2010})}\BibitemShut {NoStop}%
\bibitem [{\citenamefont {Tokura}\ \emph {et~al.}(2014)\citenamefont {Tokura},
  \citenamefont {Seki},\ and\ \citenamefont
  {Nagaosa}}]{tokura2014multiferroics}%
  \BibitemOpen
  \bibfield  {author} {\bibinfo {author} {\bibfnamefont {Y.}~\bibnamefont
  {Tokura}}, \bibinfo {author} {\bibfnamefont {S.}~\bibnamefont {Seki}}, \ and\
  \bibinfo {author} {\bibfnamefont {N.}~\bibnamefont {Nagaosa}},\ }\href
  {https://iopscience.iop.org/article/10.1088/0034-4885/77/7/076501} {\bibfield
   {journal} {\bibinfo  {journal} {Reports on Progress in Physics}\ }\textbf
  {\bibinfo {volume} {77}},\ \bibinfo {pages} {076501} (\bibinfo {year}
  {2014})}\BibitemShut {NoStop}%
\bibitem [{\citenamefont {Dong}\ \emph {et~al.}(2015)\citenamefont {Dong},
  \citenamefont {Liu}, \citenamefont {Cheong},\ and\ \citenamefont
  {Ren}}]{dong2015multiferroic}%
  \BibitemOpen
  \bibfield  {author} {\bibinfo {author} {\bibfnamefont {S.}~\bibnamefont
  {Dong}}, \bibinfo {author} {\bibfnamefont {J.-M.}\ \bibnamefont {Liu}},
  \bibinfo {author} {\bibfnamefont {S.-W.}\ \bibnamefont {Cheong}}, \ and\
  \bibinfo {author} {\bibfnamefont {Z.}~\bibnamefont {Ren}},\ }\href
  {https://www.tandfonline.com/doi/abs/10.1080/00018732.2015.1114338?journalCode=tadp20}
  {\bibfield  {journal} {\bibinfo  {journal} {Advances in Physics}\ }\textbf
  {\bibinfo {volume} {64}},\ \bibinfo {pages} {519} (\bibinfo {year}
  {2015})}\BibitemShut {NoStop}%
\bibitem [{\citenamefont {Fiebig}\ \emph {et~al.}(2016)\citenamefont {Fiebig},
  \citenamefont {Lottermoser}, \citenamefont {Meier},\ and\ \citenamefont
  {Trassin}}]{fiebig2016evolution}%
  \BibitemOpen
  \bibfield  {author} {\bibinfo {author} {\bibfnamefont {M.}~\bibnamefont
  {Fiebig}}, \bibinfo {author} {\bibfnamefont {T.}~\bibnamefont {Lottermoser}},
  \bibinfo {author} {\bibfnamefont {D.}~\bibnamefont {Meier}}, \ and\ \bibinfo
  {author} {\bibfnamefont {M.}~\bibnamefont {Trassin}},\ }\href
  {https://www.nature.com/articles/natrevmats201646} {\bibfield  {journal}
  {\bibinfo  {journal} {Nature Reviews Materials}\ }\textbf {\bibinfo {volume}
  {1}},\ \bibinfo {pages} {1} (\bibinfo {year} {2016})}\BibitemShut {NoStop}%
\bibitem [{\citenamefont {Dzyaloshinskii}(1960)}]{1571980074033746432}%
  \BibitemOpen
  \bibfield  {author} {\bibinfo {author} {\bibfnamefont {I.~E.}\ \bibnamefont
  {Dzyaloshinskii}},\ }\href {https://cir.nii.ac.jp/crid/1571980074033746432}
  {\bibfield  {journal} {\bibinfo  {journal} {Sov. Phys. JETP}\ }\textbf
  {\bibinfo {volume} {10}},\ \bibinfo {pages} {628} (\bibinfo {year}
  {1960})}\BibitemShut {NoStop}%
\bibitem [{\citenamefont {Astrov}(1960)}]{astrov1960magnetoelectric}%
  \BibitemOpen
  \bibfield  {author} {\bibinfo {author} {\bibfnamefont {D.}~\bibnamefont
  {Astrov}},\ }\href {http://www.jetp.ras.ru/cgi-bin/dn/e_011_03_0708.pdf}
  {\bibfield  {journal} {\bibinfo  {journal} {Sov. Phys. JETP}\ }\textbf
  {\bibinfo {volume} {11}},\ \bibinfo {pages} {708} (\bibinfo {year}
  {1960})}\BibitemShut {NoStop}%
\bibitem [{\citenamefont {Folen}\ \emph {et~al.}(1961)\citenamefont {Folen},
  \citenamefont {Rado},\ and\ \citenamefont {Stalder}}]{PhysRevLett.6.607}%
  \BibitemOpen
  \bibfield  {author} {\bibinfo {author} {\bibfnamefont {V.~J.}\ \bibnamefont
  {Folen}}, \bibinfo {author} {\bibfnamefont {G.~T.}\ \bibnamefont {Rado}}, \
  and\ \bibinfo {author} {\bibfnamefont {E.~W.}\ \bibnamefont {Stalder}},\
  }\href {\doibase 10.1103/PhysRevLett.6.607} {\bibfield  {journal} {\bibinfo
  {journal} {Phys. Rev. Lett.}\ }\textbf {\bibinfo {volume} {6}},\ \bibinfo
  {pages} {607} (\bibinfo {year} {1961})}\BibitemShut {NoStop}%
\bibitem [{\citenamefont {Qi}\ \emph {et~al.}(2008)\citenamefont {Qi},
  \citenamefont {Hughes},\ and\ \citenamefont {Zhang}}]{PhysRevB.78.195424}%
  \BibitemOpen
  \bibfield  {author} {\bibinfo {author} {\bibfnamefont {X.-L.}\ \bibnamefont
  {Qi}}, \bibinfo {author} {\bibfnamefont {T.~L.}\ \bibnamefont {Hughes}}, \
  and\ \bibinfo {author} {\bibfnamefont {S.-C.}\ \bibnamefont {Zhang}},\ }\href
  {\doibase 10.1103/PhysRevB.78.195424} {\bibfield  {journal} {\bibinfo
  {journal} {Phys. Rev. B}\ }\textbf {\bibinfo {volume} {78}},\ \bibinfo
  {pages} {195424} (\bibinfo {year} {2008})}\BibitemShut {NoStop}%
\bibitem [{\citenamefont {Essin}\ \emph {et~al.}(2009)\citenamefont {Essin},
  \citenamefont {Moore},\ and\ \citenamefont
  {Vanderbilt}}]{PhysRevLett.102.146805}%
  \BibitemOpen
  \bibfield  {author} {\bibinfo {author} {\bibfnamefont {A.~M.}\ \bibnamefont
  {Essin}}, \bibinfo {author} {\bibfnamefont {J.~E.}\ \bibnamefont {Moore}}, \
  and\ \bibinfo {author} {\bibfnamefont {D.}~\bibnamefont {Vanderbilt}},\
  }\href {\doibase 10.1103/PhysRevLett.102.146805} {\bibfield  {journal}
  {\bibinfo  {journal} {Phys. Rev. Lett.}\ }\textbf {\bibinfo {volume} {102}},\
  \bibinfo {pages} {146805} (\bibinfo {year} {2009})}\BibitemShut {NoStop}%
\bibitem [{\citenamefont {Malashevich}\ \emph {et~al.}(2010)\citenamefont
  {Malashevich}, \citenamefont {Souza}, \citenamefont {Coh},\ and\
  \citenamefont {Vanderbilt}}]{Malashevich_2010}%
  \BibitemOpen
  \bibfield  {author} {\bibinfo {author} {\bibfnamefont {A.}~\bibnamefont
  {Malashevich}}, \bibinfo {author} {\bibfnamefont {I.}~\bibnamefont {Souza}},
  \bibinfo {author} {\bibfnamefont {S.}~\bibnamefont {Coh}}, \ and\ \bibinfo
  {author} {\bibfnamefont {D.}~\bibnamefont {Vanderbilt}},\ }\href {\doibase
  10.1088/1367-2630/12/5/053032} {\bibfield  {journal} {\bibinfo  {journal}
  {New Journal of Physics}\ }\textbf {\bibinfo {volume} {12}},\ \bibinfo
  {pages} {053032} (\bibinfo {year} {2010})}\BibitemShut {NoStop}%
\bibitem [{\citenamefont {Malashevich}\ and\ \citenamefont
  {Souza}(2010)}]{PhysRevB.82.245118}%
  \BibitemOpen
  \bibfield  {author} {\bibinfo {author} {\bibfnamefont {A.}~\bibnamefont
  {Malashevich}}\ and\ \bibinfo {author} {\bibfnamefont {I.}~\bibnamefont
  {Souza}},\ }\href {\doibase 10.1103/PhysRevB.82.245118} {\bibfield  {journal}
  {\bibinfo  {journal} {Phys. Rev. B}\ }\textbf {\bibinfo {volume} {82}},\
  \bibinfo {pages} {245118} (\bibinfo {year} {2010})}\BibitemShut {NoStop}%
\bibitem [{\citenamefont {Gao}\ \emph {et~al.}(2014)\citenamefont {Gao},
  \citenamefont {Yang},\ and\ \citenamefont {Niu}}]{PhysRevLett.112.166601}%
  \BibitemOpen
  \bibfield  {author} {\bibinfo {author} {\bibfnamefont {Y.}~\bibnamefont
  {Gao}}, \bibinfo {author} {\bibfnamefont {S.~A.}\ \bibnamefont {Yang}}, \
  and\ \bibinfo {author} {\bibfnamefont {Q.}~\bibnamefont {Niu}},\ }\href
  {\doibase 10.1103/PhysRevLett.112.166601} {\bibfield  {journal} {\bibinfo
  {journal} {Phys. Rev. Lett.}\ }\textbf {\bibinfo {volume} {112}},\ \bibinfo
  {pages} {166601} (\bibinfo {year} {2014})}\BibitemShut {NoStop}%
\bibitem [{\citenamefont {Thonhauser}\ \emph {et~al.}(2005)\citenamefont
  {Thonhauser}, \citenamefont {Ceresoli}, \citenamefont {Vanderbilt},\ and\
  \citenamefont {Resta}}]{PhysRevLett.95.137205}%
  \BibitemOpen
  \bibfield  {author} {\bibinfo {author} {\bibfnamefont {T.}~\bibnamefont
  {Thonhauser}}, \bibinfo {author} {\bibfnamefont {D.}~\bibnamefont
  {Ceresoli}}, \bibinfo {author} {\bibfnamefont {D.}~\bibnamefont
  {Vanderbilt}}, \ and\ \bibinfo {author} {\bibfnamefont {R.}~\bibnamefont
  {Resta}},\ }\href {\doibase 10.1103/PhysRevLett.95.137205} {\bibfield
  {journal} {\bibinfo  {journal} {Phys. Rev. Lett.}\ }\textbf {\bibinfo
  {volume} {95}},\ \bibinfo {pages} {137205} (\bibinfo {year}
  {2005})}\BibitemShut {NoStop}%
\bibitem [{\citenamefont {Ceresoli}\ \emph {et~al.}(2006)\citenamefont
  {Ceresoli}, \citenamefont {Thonhauser}, \citenamefont {Vanderbilt},\ and\
  \citenamefont {Resta}}]{PhysRevB.74.024408}%
  \BibitemOpen
  \bibfield  {author} {\bibinfo {author} {\bibfnamefont {D.}~\bibnamefont
  {Ceresoli}}, \bibinfo {author} {\bibfnamefont {T.}~\bibnamefont
  {Thonhauser}}, \bibinfo {author} {\bibfnamefont {D.}~\bibnamefont
  {Vanderbilt}}, \ and\ \bibinfo {author} {\bibfnamefont {R.}~\bibnamefont
  {Resta}},\ }\href {\doibase 10.1103/PhysRevB.74.024408} {\bibfield  {journal}
  {\bibinfo  {journal} {Phys. Rev. B}\ }\textbf {\bibinfo {volume} {74}},\
  \bibinfo {pages} {024408} (\bibinfo {year} {2006})}\BibitemShut {NoStop}%
\bibitem [{\citenamefont {Xiao}\ \emph {et~al.}(2005)\citenamefont {Xiao},
  \citenamefont {Shi},\ and\ \citenamefont {Niu}}]{PhysRevLett.95.137204}%
  \BibitemOpen
  \bibfield  {author} {\bibinfo {author} {\bibfnamefont {D.}~\bibnamefont
  {Xiao}}, \bibinfo {author} {\bibfnamefont {J.}~\bibnamefont {Shi}}, \ and\
  \bibinfo {author} {\bibfnamefont {Q.}~\bibnamefont {Niu}},\ }\href {\doibase
  10.1103/PhysRevLett.95.137204} {\bibfield  {journal} {\bibinfo  {journal}
  {Phys. Rev. Lett.}\ }\textbf {\bibinfo {volume} {95}},\ \bibinfo {pages}
  {137204} (\bibinfo {year} {2005})}\BibitemShut {NoStop}%
\bibitem [{\citenamefont {Xiao}\ \emph {et~al.}(2006)\citenamefont {Xiao},
  \citenamefont {Yao}, \citenamefont {Fang},\ and\ \citenamefont
  {Niu}}]{PhysRevLett.97.026603}%
  \BibitemOpen
  \bibfield  {author} {\bibinfo {author} {\bibfnamefont {D.}~\bibnamefont
  {Xiao}}, \bibinfo {author} {\bibfnamefont {Y.}~\bibnamefont {Yao}}, \bibinfo
  {author} {\bibfnamefont {Z.}~\bibnamefont {Fang}}, \ and\ \bibinfo {author}
  {\bibfnamefont {Q.}~\bibnamefont {Niu}},\ }\href {\doibase
  10.1103/PhysRevLett.97.026603} {\bibfield  {journal} {\bibinfo  {journal}
  {Phys. Rev. Lett.}\ }\textbf {\bibinfo {volume} {97}},\ \bibinfo {pages}
  {026603} (\bibinfo {year} {2006})}\BibitemShut {NoStop}%
\bibitem [{\citenamefont {Shi}\ \emph {et~al.}(2007)\citenamefont {Shi},
  \citenamefont {Vignale}, \citenamefont {Xiao},\ and\ \citenamefont
  {Niu}}]{PhysRevLett.99.197202}%
  \BibitemOpen
  \bibfield  {author} {\bibinfo {author} {\bibfnamefont {J.}~\bibnamefont
  {Shi}}, \bibinfo {author} {\bibfnamefont {G.}~\bibnamefont {Vignale}},
  \bibinfo {author} {\bibfnamefont {D.}~\bibnamefont {Xiao}}, \ and\ \bibinfo
  {author} {\bibfnamefont {Q.}~\bibnamefont {Niu}},\ }\href {\doibase
  10.1103/PhysRevLett.99.197202} {\bibfield  {journal} {\bibinfo  {journal}
  {Phys. Rev. Lett.}\ }\textbf {\bibinfo {volume} {99}},\ \bibinfo {pages}
  {197202} (\bibinfo {year} {2007})}\BibitemShut {NoStop}%
\bibitem [{\citenamefont {Lee}\ \emph {et~al.}(2017)\citenamefont {Lee},
  \citenamefont {Wang}, \citenamefont {Xie}, \citenamefont {Mak},\ and\
  \citenamefont {Shan}}]{lee2017valley}%
  \BibitemOpen
  \bibfield  {author} {\bibinfo {author} {\bibfnamefont {J.}~\bibnamefont
  {Lee}}, \bibinfo {author} {\bibfnamefont {Z.}~\bibnamefont {Wang}}, \bibinfo
  {author} {\bibfnamefont {H.}~\bibnamefont {Xie}}, \bibinfo {author}
  {\bibfnamefont {K.~F.}\ \bibnamefont {Mak}}, \ and\ \bibinfo {author}
  {\bibfnamefont {J.}~\bibnamefont {Shan}},\ }\href
  {https://www.nature.com/articles/nmat4931} {\bibfield  {journal} {\bibinfo
  {journal} {Nature materials}\ }\textbf {\bibinfo {volume} {16}},\ \bibinfo
  {pages} {887} (\bibinfo {year} {2017})}\BibitemShut {NoStop}%
\bibitem [{\citenamefont {Son}\ \emph {et~al.}(2019)\citenamefont {Son},
  \citenamefont {Kim}, \citenamefont {Ahn}, \citenamefont {Lee},\ and\
  \citenamefont {Lee}}]{PhysRevLett.123.036806}%
  \BibitemOpen
  \bibfield  {author} {\bibinfo {author} {\bibfnamefont {J.}~\bibnamefont
  {Son}}, \bibinfo {author} {\bibfnamefont {K.-H.}\ \bibnamefont {Kim}},
  \bibinfo {author} {\bibfnamefont {Y.~H.}\ \bibnamefont {Ahn}}, \bibinfo
  {author} {\bibfnamefont {H.-W.}\ \bibnamefont {Lee}}, \ and\ \bibinfo
  {author} {\bibfnamefont {J.}~\bibnamefont {Lee}},\ }\href {\doibase
  10.1103/PhysRevLett.123.036806} {\bibfield  {journal} {\bibinfo  {journal}
  {Phys. Rev. Lett.}\ }\textbf {\bibinfo {volume} {123}},\ \bibinfo {pages}
  {036806} (\bibinfo {year} {2019})}\BibitemShut {NoStop}%
\bibitem [{\citenamefont {Sharpe}\ \emph {et~al.}(2019)\citenamefont {Sharpe},
  \citenamefont {Fox}, \citenamefont {Barnard}, \citenamefont {Finney},
  \citenamefont {Watanabe}, \citenamefont {Taniguchi}, \citenamefont
  {Kastner},\ and\ \citenamefont
  {Goldhaber-Gordon}}]{doi:10.1126/science.aaw3780}%
  \BibitemOpen
  \bibfield  {author} {\bibinfo {author} {\bibfnamefont {A.~L.}\ \bibnamefont
  {Sharpe}}, \bibinfo {author} {\bibfnamefont {E.~J.}\ \bibnamefont {Fox}},
  \bibinfo {author} {\bibfnamefont {A.~W.}\ \bibnamefont {Barnard}}, \bibinfo
  {author} {\bibfnamefont {J.}~\bibnamefont {Finney}}, \bibinfo {author}
  {\bibfnamefont {K.}~\bibnamefont {Watanabe}}, \bibinfo {author}
  {\bibfnamefont {T.}~\bibnamefont {Taniguchi}}, \bibinfo {author}
  {\bibfnamefont {M.~A.}\ \bibnamefont {Kastner}}, \ and\ \bibinfo {author}
  {\bibfnamefont {D.}~\bibnamefont {Goldhaber-Gordon}},\ }\href {\doibase
  10.1126/science.aaw3780} {\bibfield  {journal} {\bibinfo  {journal}
  {Science}\ }\textbf {\bibinfo {volume} {365}},\ \bibinfo {pages} {605}
  (\bibinfo {year} {2019})}\BibitemShut {NoStop}%
\bibitem [{\citenamefont {He}\ \emph {et~al.}(2020)\citenamefont {He},
  \citenamefont {Goldhaber-Gordon},\ and\ \citenamefont {Law}}]{he2020giant}%
  \BibitemOpen
  \bibfield  {author} {\bibinfo {author} {\bibfnamefont {W.-Y.}\ \bibnamefont
  {He}}, \bibinfo {author} {\bibfnamefont {D.}~\bibnamefont
  {Goldhaber-Gordon}}, \ and\ \bibinfo {author} {\bibfnamefont {K.~T.}\
  \bibnamefont {Law}},\ }\href
  {https://www.nature.com/articles/s41467-020-15473-9} {\bibfield  {journal}
  {\bibinfo  {journal} {Nature communications}\ }\textbf {\bibinfo {volume}
  {11}},\ \bibinfo {pages} {1} (\bibinfo {year} {2020})}\BibitemShut {NoStop}%
\bibitem [{\citenamefont {Yoda}\ \emph {et~al.}(2015)\citenamefont {Yoda},
  \citenamefont {Yokoyama},\ and\ \citenamefont {Murakami}}]{yoda2015current}%
  \BibitemOpen
  \bibfield  {author} {\bibinfo {author} {\bibfnamefont {T.}~\bibnamefont
  {Yoda}}, \bibinfo {author} {\bibfnamefont {T.}~\bibnamefont {Yokoyama}}, \
  and\ \bibinfo {author} {\bibfnamefont {S.}~\bibnamefont {Murakami}},\ }\href
  {https://www.nature.com/articles/srep12024} {\bibfield  {journal} {\bibinfo
  {journal} {Scientific reports}\ }\textbf {\bibinfo {volume} {5}},\ \bibinfo
  {pages} {1} (\bibinfo {year} {2015})}\BibitemShut {NoStop}%
\bibitem [{\citenamefont {Zhong}\ \emph {et~al.}(2016)\citenamefont {Zhong},
  \citenamefont {Moore},\ and\ \citenamefont {Souza}}]{PhysRevLett.116.077201}%
  \BibitemOpen
  \bibfield  {author} {\bibinfo {author} {\bibfnamefont {S.}~\bibnamefont
  {Zhong}}, \bibinfo {author} {\bibfnamefont {J.~E.}\ \bibnamefont {Moore}}, \
  and\ \bibinfo {author} {\bibfnamefont {I.}~\bibnamefont {Souza}},\ }\href
  {\doibase 10.1103/PhysRevLett.116.077201} {\bibfield  {journal} {\bibinfo
  {journal} {Phys. Rev. Lett.}\ }\textbf {\bibinfo {volume} {116}},\ \bibinfo
  {pages} {077201} (\bibinfo {year} {2016})}\BibitemShut {NoStop}%
\bibitem [{\citenamefont {Hara}\ \emph {et~al.}(2020)\citenamefont {Hara},
  \citenamefont {Bahramy},\ and\ \citenamefont
  {Murakami}}]{PhysRevB.102.184404}%
  \BibitemOpen
  \bibfield  {author} {\bibinfo {author} {\bibfnamefont {D.}~\bibnamefont
  {Hara}}, \bibinfo {author} {\bibfnamefont {M.~S.}\ \bibnamefont {Bahramy}}, \
  and\ \bibinfo {author} {\bibfnamefont {S.}~\bibnamefont {Murakami}},\ }\href
  {\doibase 10.1103/PhysRevB.102.184404} {\bibfield  {journal} {\bibinfo
  {journal} {Phys. Rev. B}\ }\textbf {\bibinfo {volume} {102}},\ \bibinfo
  {pages} {184404} (\bibinfo {year} {2020})}\BibitemShut {NoStop}%
\bibitem [{\citenamefont {Bhowal}\ and\ \citenamefont
  {Satpathy}(2020)}]{PhysRevB.102.201403}%
  \BibitemOpen
  \bibfield  {author} {\bibinfo {author} {\bibfnamefont {S.}~\bibnamefont
  {Bhowal}}\ and\ \bibinfo {author} {\bibfnamefont {S.}~\bibnamefont
  {Satpathy}},\ }\href {\doibase 10.1103/PhysRevB.102.201403} {\bibfield
  {journal} {\bibinfo  {journal} {Phys. Rev. B}\ }\textbf {\bibinfo {volume}
  {102}},\ \bibinfo {pages} {201403} (\bibinfo {year} {2020})}\BibitemShut
  {NoStop}%
\bibitem [{\citenamefont {Rou}\ \emph {et~al.}(2017)\citenamefont {Rou},
  \citenamefont {\ifmmode~\mbox{\c{S}}\else \c{S}\fi{}ahin}, \citenamefont
  {Ma},\ and\ \citenamefont {Pesin}}]{PhysRevB.96.035120}%
  \BibitemOpen
  \bibfield  {author} {\bibinfo {author} {\bibfnamefont {J.}~\bibnamefont
  {Rou}}, \bibinfo {author} {\bibfnamefont {C.}~\bibnamefont
  {\ifmmode~\mbox{\c{S}}\else \c{S}\fi{}ahin}}, \bibinfo {author}
  {\bibfnamefont {J.}~\bibnamefont {Ma}}, \ and\ \bibinfo {author}
  {\bibfnamefont {D.~A.}\ \bibnamefont {Pesin}},\ }\href {\doibase
  10.1103/PhysRevB.96.035120} {\bibfield  {journal} {\bibinfo  {journal} {Phys.
  Rev. B}\ }\textbf {\bibinfo {volume} {96}},\ \bibinfo {pages} {035120}
  (\bibinfo {year} {2017})}\BibitemShut {NoStop}%
\bibitem [{\citenamefont {He}\ and\ \citenamefont
  {Law}(2021)}]{PhysRevResearch.3.L032012}%
  \BibitemOpen
  \bibfield  {author} {\bibinfo {author} {\bibfnamefont {W.-Y.}\ \bibnamefont
  {He}}\ and\ \bibinfo {author} {\bibfnamefont {K.~T.}\ \bibnamefont {Law}},\
  }\href {\doibase 10.1103/PhysRevResearch.3.L032012} {\bibfield  {journal}
  {\bibinfo  {journal} {Phys. Rev. Research}\ }\textbf {\bibinfo {volume}
  {3}},\ \bibinfo {pages} {L032012} (\bibinfo {year} {2021})}\BibitemShut
  {NoStop}%
\bibitem [{\citenamefont {Chirolli}\ \emph {et~al.}(2022)\citenamefont
  {Chirolli}, \citenamefont {Mercaldo}, \citenamefont {Guarcello},
  \citenamefont {Giazotto},\ and\ \citenamefont
  {Cuoco}}]{PhysRevLett.128.217703}%
  \BibitemOpen
  \bibfield  {author} {\bibinfo {author} {\bibfnamefont {L.}~\bibnamefont
  {Chirolli}}, \bibinfo {author} {\bibfnamefont {M.~T.}\ \bibnamefont
  {Mercaldo}}, \bibinfo {author} {\bibfnamefont {C.}~\bibnamefont {Guarcello}},
  \bibinfo {author} {\bibfnamefont {F.}~\bibnamefont {Giazotto}}, \ and\
  \bibinfo {author} {\bibfnamefont {M.}~\bibnamefont {Cuoco}},\ }\href
  {\doibase 10.1103/PhysRevLett.128.217703} {\bibfield  {journal} {\bibinfo
  {journal} {Phys. Rev. Lett.}\ }\textbf {\bibinfo {volume} {128}},\ \bibinfo
  {pages} {217703} (\bibinfo {year} {2022})}\BibitemShut {NoStop}%
\bibitem [{\citenamefont {Xiao}\ \emph
  {et~al.}(2021{\natexlab{a}})\citenamefont {Xiao}, \citenamefont {Ren},\ and\
  \citenamefont {Xiong}}]{PhysRevB.103.115432}%
  \BibitemOpen
  \bibfield  {author} {\bibinfo {author} {\bibfnamefont {C.}~\bibnamefont
  {Xiao}}, \bibinfo {author} {\bibfnamefont {Y.}~\bibnamefont {Ren}}, \ and\
  \bibinfo {author} {\bibfnamefont {B.}~\bibnamefont {Xiong}},\ }\href
  {\doibase 10.1103/PhysRevB.103.115432} {\bibfield  {journal} {\bibinfo
  {journal} {Phys. Rev. B}\ }\textbf {\bibinfo {volume} {103}},\ \bibinfo
  {pages} {115432} (\bibinfo {year} {2021}{\natexlab{a}})}\BibitemShut
  {NoStop}%
\bibitem [{\citenamefont {Xiao}\ \emph
  {et~al.}(2021{\natexlab{b}})\citenamefont {Xiao}, \citenamefont {Liu},
  \citenamefont {Zhao}, \citenamefont {Yang},\ and\ \citenamefont
  {Niu}}]{PhysRevB.103.045401}%
  \BibitemOpen
  \bibfield  {author} {\bibinfo {author} {\bibfnamefont {C.}~\bibnamefont
  {Xiao}}, \bibinfo {author} {\bibfnamefont {H.}~\bibnamefont {Liu}}, \bibinfo
  {author} {\bibfnamefont {J.}~\bibnamefont {Zhao}}, \bibinfo {author}
  {\bibfnamefont {S.~A.}\ \bibnamefont {Yang}}, \ and\ \bibinfo {author}
  {\bibfnamefont {Q.}~\bibnamefont {Niu}},\ }\href {\doibase
  10.1103/PhysRevB.103.045401} {\bibfield  {journal} {\bibinfo  {journal}
  {Phys. Rev. B}\ }\textbf {\bibinfo {volume} {103}},\ \bibinfo {pages}
  {045401} (\bibinfo {year} {2021}{\natexlab{b}})}\BibitemShut {NoStop}%
\bibitem [{\citenamefont {Gao}\ and\ \citenamefont
  {Xiao}(2019)}]{PhysRevLett.122.227402}%
  \BibitemOpen
  \bibfield  {author} {\bibinfo {author} {\bibfnamefont {Y.}~\bibnamefont
  {Gao}}\ and\ \bibinfo {author} {\bibfnamefont {D.}~\bibnamefont {Xiao}},\
  }\href {\doibase 10.1103/PhysRevLett.122.227402} {\bibfield  {journal}
  {\bibinfo  {journal} {Phys. Rev. Lett.}\ }\textbf {\bibinfo {volume} {122}},\
  \bibinfo {pages} {227402} (\bibinfo {year} {2019})}\BibitemShut {NoStop}%
\bibitem [{\citenamefont {Altland}\ and\ \citenamefont
  {Simons}(2010)}]{altland_simons_2010}%
  \BibitemOpen
  \bibfield  {author} {\bibinfo {author} {\bibfnamefont {A.}~\bibnamefont
  {Altland}}\ and\ \bibinfo {author} {\bibfnamefont {B.~D.}\ \bibnamefont
  {Simons}},\ }\href {\doibase 10.1017/CBO9780511789984} {\emph {\bibinfo
  {title} {Condensed Matter Field Theory}}},\ \bibinfo {edition} {2nd}\ ed.\
  (\bibinfo  {publisher} {Cambridge University Press},\ \bibinfo {year}
  {2010})\BibitemShut {NoStop}%
\bibitem [{\citenamefont {Raab}\ and\ \citenamefont
  {de~Lange}(2004)}]{10.1093/acprof:oso/9780198567271.001.0001}%
  \BibitemOpen
  \bibfield  {author} {\bibinfo {author} {\bibfnamefont {R.~E.}\ \bibnamefont
  {Raab}}\ and\ \bibinfo {author} {\bibfnamefont {O.~L.}\ \bibnamefont
  {de~Lange}},\ }\href {\doibase 10.1093/acprof:oso/9780198567271.001.0001}
  {\emph {\bibinfo {title} {{Multipole Theory in Electromagnetism: Classical,
  quantum, and symmetry aspects, with applications}}}}\ (\bibinfo  {publisher}
  {Oxford University Press},\ \bibinfo {year} {2004})\BibitemShut {NoStop}%
\bibitem [{\citenamefont {Provost}\ and\ \citenamefont
  {Vallee}(1980)}]{provost1980riemannian}%
  \BibitemOpen
  \bibfield  {author} {\bibinfo {author} {\bibfnamefont {J.}~\bibnamefont
  {Provost}}\ and\ \bibinfo {author} {\bibfnamefont {G.}~\bibnamefont
  {Vallee}},\ }\href {https://link.springer.com/article/10.1007/BF02193559}
  {\bibfield  {journal} {\bibinfo  {journal} {Communications in Mathematical
  Physics}\ }\textbf {\bibinfo {volume} {76}},\ \bibinfo {pages} {289}
  (\bibinfo {year} {1980})}\BibitemShut {NoStop}%
\bibitem [{\citenamefont {Resta}(2011)}]{resta2011insulating}%
  \BibitemOpen
  \bibfield  {author} {\bibinfo {author} {\bibfnamefont {R.}~\bibnamefont
  {Resta}},\ }\href
  {https://link.springer.com/article/10.1140/epjb/e2010-10874-4} {\bibfield
  {journal} {\bibinfo  {journal} {The European Physical Journal B}\ }\textbf
  {\bibinfo {volume} {79}},\ \bibinfo {pages} {121} (\bibinfo {year}
  {2011})}\BibitemShut {NoStop}%
\bibitem [{\citenamefont {Lapa}\ and\ \citenamefont
  {Hughes}(2019)}]{PhysRevB.99.121111}%
  \BibitemOpen
  \bibfield  {author} {\bibinfo {author} {\bibfnamefont {M.~F.}\ \bibnamefont
  {Lapa}}\ and\ \bibinfo {author} {\bibfnamefont {T.~L.}\ \bibnamefont
  {Hughes}},\ }\href {\doibase 10.1103/PhysRevB.99.121111} {\bibfield
  {journal} {\bibinfo  {journal} {Phys. Rev. B}\ }\textbf {\bibinfo {volume}
  {99}},\ \bibinfo {pages} {121111} (\bibinfo {year} {2019})}\BibitemShut
  {NoStop}%
\bibitem [{\citenamefont {Daido}\ \emph {et~al.}(2020)\citenamefont {Daido},
  \citenamefont {Shitade},\ and\ \citenamefont {Yanase}}]{PhysRevB.102.235149}%
  \BibitemOpen
  \bibfield  {author} {\bibinfo {author} {\bibfnamefont {A.}~\bibnamefont
  {Daido}}, \bibinfo {author} {\bibfnamefont {A.}~\bibnamefont {Shitade}}, \
  and\ \bibinfo {author} {\bibfnamefont {Y.}~\bibnamefont {Yanase}},\ }\href
  {\doibase 10.1103/PhysRevB.102.235149} {\bibfield  {journal} {\bibinfo
  {journal} {Phys. Rev. B}\ }\textbf {\bibinfo {volume} {102}},\ \bibinfo
  {pages} {235149} (\bibinfo {year} {2020})}\BibitemShut {NoStop}%
\bibitem [{\citenamefont {Shitade}\ \emph {et~al.}(2018)\citenamefont
  {Shitade}, \citenamefont {Watanabe},\ and\ \citenamefont
  {Yanase}}]{PhysRevB.98.020407}%
  \BibitemOpen
  \bibfield  {author} {\bibinfo {author} {\bibfnamefont {A.}~\bibnamefont
  {Shitade}}, \bibinfo {author} {\bibfnamefont {H.}~\bibnamefont {Watanabe}}, \
  and\ \bibinfo {author} {\bibfnamefont {Y.}~\bibnamefont {Yanase}},\ }\href
  {\doibase 10.1103/PhysRevB.98.020407} {\bibfield  {journal} {\bibinfo
  {journal} {Phys. Rev. B}\ }\textbf {\bibinfo {volume} {98}},\ \bibinfo
  {pages} {020407} (\bibinfo {year} {2018})}\BibitemShut {NoStop}%
\bibitem [{\citenamefont {Gao}\ and\ \citenamefont
  {Xiao}(2018)}]{PhysRevB.98.060402}%
  \BibitemOpen
  \bibfield  {author} {\bibinfo {author} {\bibfnamefont {Y.}~\bibnamefont
  {Gao}}\ and\ \bibinfo {author} {\bibfnamefont {D.}~\bibnamefont {Xiao}},\
  }\href {\doibase 10.1103/PhysRevB.98.060402} {\bibfield  {journal} {\bibinfo
  {journal} {Phys. Rev. B}\ }\textbf {\bibinfo {volume} {98}},\ \bibinfo
  {pages} {060402} (\bibinfo {year} {2018})}\BibitemShut {NoStop}%
\bibitem [{\citenamefont {Ederer}\ and\ \citenamefont
  {Spaldin}(2007)}]{PhysRevB.76.214404}%
  \BibitemOpen
  \bibfield  {author} {\bibinfo {author} {\bibfnamefont {C.}~\bibnamefont
  {Ederer}}\ and\ \bibinfo {author} {\bibfnamefont {N.~A.}\ \bibnamefont
  {Spaldin}},\ }\href {\doibase 10.1103/PhysRevB.76.214404} {\bibfield
  {journal} {\bibinfo  {journal} {Phys. Rev. B}\ }\textbf {\bibinfo {volume}
  {76}},\ \bibinfo {pages} {214404} (\bibinfo {year} {2007})}\BibitemShut
  {NoStop}%
\bibitem [{\citenamefont {Spaldin}\ \emph {et~al.}(2008)\citenamefont
  {Spaldin}, \citenamefont {Fiebig},\ and\ \citenamefont
  {Mostovoy}}]{spaldin2008toroidal}%
  \BibitemOpen
  \bibfield  {author} {\bibinfo {author} {\bibfnamefont {N.~A.}\ \bibnamefont
  {Spaldin}}, \bibinfo {author} {\bibfnamefont {M.}~\bibnamefont {Fiebig}}, \
  and\ \bibinfo {author} {\bibfnamefont {M.}~\bibnamefont {Mostovoy}},\ }\href
  {https://iopscience.iop.org/article/10.1088/0953-8984/20/43/434203}
  {\bibfield  {journal} {\bibinfo  {journal} {Journal of Physics: Condensed
  Matter}\ }\textbf {\bibinfo {volume} {20}},\ \bibinfo {pages} {434203}
  (\bibinfo {year} {2008})}\BibitemShut {NoStop}%
\bibitem [{\citenamefont {Spaldin}\ \emph {et~al.}(2013)\citenamefont
  {Spaldin}, \citenamefont {Fechner}, \citenamefont {Bousquet}, \citenamefont
  {Balatsky},\ and\ \citenamefont {Nordstr\"om}}]{PhysRevB.88.094429}%
  \BibitemOpen
  \bibfield  {author} {\bibinfo {author} {\bibfnamefont {N.~A.}\ \bibnamefont
  {Spaldin}}, \bibinfo {author} {\bibfnamefont {M.}~\bibnamefont {Fechner}},
  \bibinfo {author} {\bibfnamefont {E.}~\bibnamefont {Bousquet}}, \bibinfo
  {author} {\bibfnamefont {A.}~\bibnamefont {Balatsky}}, \ and\ \bibinfo
  {author} {\bibfnamefont {L.}~\bibnamefont {Nordstr\"om}},\ }\href {\doibase
  10.1103/PhysRevB.88.094429} {\bibfield  {journal} {\bibinfo  {journal} {Phys.
  Rev. B}\ }\textbf {\bibinfo {volume} {88}},\ \bibinfo {pages} {094429}
  (\bibinfo {year} {2013})}\BibitemShut {NoStop}%
\bibitem [{\citenamefont {Th\"ole}\ \emph {et~al.}(2016)\citenamefont
  {Th\"ole}, \citenamefont {Fechner},\ and\ \citenamefont
  {Spaldin}}]{PhysRevB.93.195167}%
  \BibitemOpen
  \bibfield  {author} {\bibinfo {author} {\bibfnamefont {F.}~\bibnamefont
  {Th\"ole}}, \bibinfo {author} {\bibfnamefont {M.}~\bibnamefont {Fechner}}, \
  and\ \bibinfo {author} {\bibfnamefont {N.~A.}\ \bibnamefont {Spaldin}},\
  }\href {\doibase 10.1103/PhysRevB.93.195167} {\bibfield  {journal} {\bibinfo
  {journal} {Phys. Rev. B}\ }\textbf {\bibinfo {volume} {93}},\ \bibinfo
  {pages} {195167} (\bibinfo {year} {2016})}\BibitemShut {NoStop}%
\bibitem [{\citenamefont {Streda}(1982)}]{Streda_1982}%
  \BibitemOpen
  \bibfield  {author} {\bibinfo {author} {\bibfnamefont {P.}~\bibnamefont
  {Streda}},\ }\href {\doibase 10.1088/0022-3719/15/22/005} {\bibfield
  {journal} {\bibinfo  {journal} {Journal of Physics C: Solid State Physics}\
  }\textbf {\bibinfo {volume} {15}},\ \bibinfo {pages} {L717} (\bibinfo {year}
  {1982})}\BibitemShut {NoStop}%
\bibitem [{\citenamefont {Winkler}\ and\ \citenamefont
  {Z\"ulicke}(2020)}]{PhysRevResearch.2.043060}%
  \BibitemOpen
  \bibfield  {author} {\bibinfo {author} {\bibfnamefont {R.}~\bibnamefont
  {Winkler}}\ and\ \bibinfo {author} {\bibfnamefont {U.}~\bibnamefont
  {Z\"ulicke}},\ }\href {\doibase 10.1103/PhysRevResearch.2.043060} {\bibfield
  {journal} {\bibinfo  {journal} {Phys. Rev. Res.}\ }\textbf {\bibinfo {volume}
  {2}},\ \bibinfo {pages} {043060} (\bibinfo {year} {2020})}\BibitemShut
  {NoStop}%
\bibitem [{\citenamefont {Varma}(1997)}]{PhysRevB.55.14554}%
  \BibitemOpen
  \bibfield  {author} {\bibinfo {author} {\bibfnamefont {C.~M.}\ \bibnamefont
  {Varma}},\ }\href {\doibase 10.1103/PhysRevB.55.14554} {\bibfield  {journal}
  {\bibinfo  {journal} {Phys. Rev. B}\ }\textbf {\bibinfo {volume} {55}},\
  \bibinfo {pages} {14554} (\bibinfo {year} {1997})}\BibitemShut {NoStop}%
\bibitem [{\citenamefont {Fauqu\'e}\ \emph {et~al.}(2006)\citenamefont
  {Fauqu\'e}, \citenamefont {Sidis}, \citenamefont {Hinkov}, \citenamefont
  {Pailh\`es}, \citenamefont {Lin}, \citenamefont {Chaud},\ and\ \citenamefont
  {Bourges}}]{PhysRevLett.96.197001}%
  \BibitemOpen
  \bibfield  {author} {\bibinfo {author} {\bibfnamefont {B.}~\bibnamefont
  {Fauqu\'e}}, \bibinfo {author} {\bibfnamefont {Y.}~\bibnamefont {Sidis}},
  \bibinfo {author} {\bibfnamefont {V.}~\bibnamefont {Hinkov}}, \bibinfo
  {author} {\bibfnamefont {S.}~\bibnamefont {Pailh\`es}}, \bibinfo {author}
  {\bibfnamefont {C.~T.}\ \bibnamefont {Lin}}, \bibinfo {author} {\bibfnamefont
  {X.}~\bibnamefont {Chaud}}, \ and\ \bibinfo {author} {\bibfnamefont
  {P.}~\bibnamefont {Bourges}},\ }\href {\doibase
  10.1103/PhysRevLett.96.197001} {\bibfield  {journal} {\bibinfo  {journal}
  {Phys. Rev. Lett.}\ }\textbf {\bibinfo {volume} {96}},\ \bibinfo {pages}
  {197001} (\bibinfo {year} {2006})}\BibitemShut {NoStop}%
\bibitem [{\citenamefont {Li}\ \emph {et~al.}(2008)\citenamefont {Li},
  \citenamefont {Bal{\'e}dent}, \citenamefont {Bari{\v{s}}i{\'c}},
  \citenamefont {Cho}, \citenamefont {Fauqu{\'e}}, \citenamefont {Sidis},
  \citenamefont {Yu}, \citenamefont {Zhao}, \citenamefont {Bourges},\ and\
  \citenamefont {Greven}}]{li2008unusual}%
  \BibitemOpen
  \bibfield  {author} {\bibinfo {author} {\bibfnamefont {Y.}~\bibnamefont
  {Li}}, \bibinfo {author} {\bibfnamefont {V.}~\bibnamefont {Bal{\'e}dent}},
  \bibinfo {author} {\bibfnamefont {N.}~\bibnamefont {Bari{\v{s}}i{\'c}}},
  \bibinfo {author} {\bibfnamefont {Y.}~\bibnamefont {Cho}}, \bibinfo {author}
  {\bibfnamefont {B.}~\bibnamefont {Fauqu{\'e}}}, \bibinfo {author}
  {\bibfnamefont {Y.}~\bibnamefont {Sidis}}, \bibinfo {author} {\bibfnamefont
  {G.}~\bibnamefont {Yu}}, \bibinfo {author} {\bibfnamefont {X.}~\bibnamefont
  {Zhao}}, \bibinfo {author} {\bibfnamefont {P.}~\bibnamefont {Bourges}}, \
  and\ \bibinfo {author} {\bibfnamefont {M.}~\bibnamefont {Greven}},\ }\href
  {https://www.nature.com/articles/nature07251} {\bibfield  {journal} {\bibinfo
   {journal} {Nature}\ }\textbf {\bibinfo {volume} {455}},\ \bibinfo {pages}
  {372} (\bibinfo {year} {2008})}\BibitemShut {NoStop}%
\bibitem [{\citenamefont {Bourges}\ and\ \citenamefont
  {Sidis}(2011)}]{BOURGES2011461}%
  \BibitemOpen
  \bibfield  {author} {\bibinfo {author} {\bibfnamefont {P.}~\bibnamefont
  {Bourges}}\ and\ \bibinfo {author} {\bibfnamefont {Y.}~\bibnamefont
  {Sidis}},\ }\href {\doibase https://doi.org/10.1016/j.crhy.2011.04.006}
  {\bibfield  {journal} {\bibinfo  {journal} {Comptes Rendus Physique}\
  }\textbf {\bibinfo {volume} {12}},\ \bibinfo {pages} {461} (\bibinfo {year}
  {2011})}\BibitemShut {NoStop}%
\bibitem [{\citenamefont {Pershoguba}\ \emph {et~al.}(2013)\citenamefont
  {Pershoguba}, \citenamefont {Kechedzhi},\ and\ \citenamefont
  {Yakovenko}}]{PhysRevLett.111.047005}%
  \BibitemOpen
  \bibfield  {author} {\bibinfo {author} {\bibfnamefont {S.~S.}\ \bibnamefont
  {Pershoguba}}, \bibinfo {author} {\bibfnamefont {K.}~\bibnamefont
  {Kechedzhi}}, \ and\ \bibinfo {author} {\bibfnamefont {V.~M.}\ \bibnamefont
  {Yakovenko}},\ }\href {\doibase 10.1103/PhysRevLett.111.047005} {\bibfield
  {journal} {\bibinfo  {journal} {Phys. Rev. Lett.}\ }\textbf {\bibinfo
  {volume} {111}},\ \bibinfo {pages} {047005} (\bibinfo {year}
  {2013})}\BibitemShut {NoStop}%
\bibitem [{\citenamefont {Zhao}\ \emph {et~al.}(2017)\citenamefont {Zhao},
  \citenamefont {Belvin}, \citenamefont {Liang}, \citenamefont {Bonn},
  \citenamefont {Hardy}, \citenamefont {Armitage},\ and\ \citenamefont
  {Hsieh}}]{zhao2017global}%
  \BibitemOpen
  \bibfield  {author} {\bibinfo {author} {\bibfnamefont {L.}~\bibnamefont
  {Zhao}}, \bibinfo {author} {\bibfnamefont {C.}~\bibnamefont {Belvin}},
  \bibinfo {author} {\bibfnamefont {R.}~\bibnamefont {Liang}}, \bibinfo
  {author} {\bibfnamefont {D.}~\bibnamefont {Bonn}}, \bibinfo {author}
  {\bibfnamefont {W.}~\bibnamefont {Hardy}}, \bibinfo {author} {\bibfnamefont
  {N.}~\bibnamefont {Armitage}}, \ and\ \bibinfo {author} {\bibfnamefont
  {D.}~\bibnamefont {Hsieh}},\ }\href
  {https://www.nature.com/articles/nphys3962} {\bibfield  {journal} {\bibinfo
  {journal} {Nature Physics}\ }\textbf {\bibinfo {volume} {13}},\ \bibinfo
  {pages} {250} (\bibinfo {year} {2017})}\BibitemShut {NoStop}%
\bibitem [{\citenamefont {Mielke}\ \emph {et~al.}(2022)\citenamefont {Mielke},
  \citenamefont {Das}, \citenamefont {Yin}, \citenamefont {Liu}, \citenamefont
  {Gupta}, \citenamefont {Jiang}, \citenamefont {Medarde}, \citenamefont {Wu},
  \citenamefont {Lei}, \citenamefont {Chang} \emph {et~al.}}]{mielke2022time}%
  \BibitemOpen
  \bibfield  {author} {\bibinfo {author} {\bibfnamefont {C.}~\bibnamefont
  {Mielke}}, \bibinfo {author} {\bibfnamefont {D.}~\bibnamefont {Das}},
  \bibinfo {author} {\bibfnamefont {J.-X.}\ \bibnamefont {Yin}}, \bibinfo
  {author} {\bibfnamefont {H.}~\bibnamefont {Liu}}, \bibinfo {author}
  {\bibfnamefont {R.}~\bibnamefont {Gupta}}, \bibinfo {author} {\bibfnamefont
  {Y.-X.}\ \bibnamefont {Jiang}}, \bibinfo {author} {\bibfnamefont
  {M.}~\bibnamefont {Medarde}}, \bibinfo {author} {\bibfnamefont
  {X.}~\bibnamefont {Wu}}, \bibinfo {author} {\bibfnamefont {H.}~\bibnamefont
  {Lei}}, \bibinfo {author} {\bibfnamefont {J.}~\bibnamefont {Chang}},  \emph
  {et~al.},\ }\href {https://www.nature.com/articles/s41586-021-04327-z}
  {\bibfield  {journal} {\bibinfo  {journal} {Nature}\ }\textbf {\bibinfo
  {volume} {602}},\ \bibinfo {pages} {245} (\bibinfo {year}
  {2022})}\BibitemShut {NoStop}%
\bibitem [{\citenamefont {Zhao}\ \emph {et~al.}(2016)\citenamefont {Zhao},
  \citenamefont {Torchinsky}, \citenamefont {Chu}, \citenamefont {Ivanov},
  \citenamefont {Lifshitz}, \citenamefont {Flint}, \citenamefont {Qi},
  \citenamefont {Cao},\ and\ \citenamefont {Hsieh}}]{zhao2016evidence}%
  \BibitemOpen
  \bibfield  {author} {\bibinfo {author} {\bibfnamefont {L.}~\bibnamefont
  {Zhao}}, \bibinfo {author} {\bibfnamefont {D.}~\bibnamefont {Torchinsky}},
  \bibinfo {author} {\bibfnamefont {H.}~\bibnamefont {Chu}}, \bibinfo {author}
  {\bibfnamefont {V.}~\bibnamefont {Ivanov}}, \bibinfo {author} {\bibfnamefont
  {R.}~\bibnamefont {Lifshitz}}, \bibinfo {author} {\bibfnamefont
  {R.}~\bibnamefont {Flint}}, \bibinfo {author} {\bibfnamefont
  {T.}~\bibnamefont {Qi}}, \bibinfo {author} {\bibfnamefont {G.}~\bibnamefont
  {Cao}}, \ and\ \bibinfo {author} {\bibfnamefont {D.}~\bibnamefont {Hsieh}},\
  }\href {https://www.nature.com/articles/nphys3517} {\bibfield  {journal}
  {\bibinfo  {journal} {Nature Physics}\ }\textbf {\bibinfo {volume} {12}},\
  \bibinfo {pages} {32} (\bibinfo {year} {2016})}\BibitemShut {NoStop}%
\bibitem [{\citenamefont {Jeong}\ \emph {et~al.}(2017)\citenamefont {Jeong},
  \citenamefont {Sidis}, \citenamefont {Louat}, \citenamefont {Brouet},\ and\
  \citenamefont {Bourges}}]{jeong2017time}%
  \BibitemOpen
  \bibfield  {author} {\bibinfo {author} {\bibfnamefont {J.}~\bibnamefont
  {Jeong}}, \bibinfo {author} {\bibfnamefont {Y.}~\bibnamefont {Sidis}},
  \bibinfo {author} {\bibfnamefont {A.}~\bibnamefont {Louat}}, \bibinfo
  {author} {\bibfnamefont {V.}~\bibnamefont {Brouet}}, \ and\ \bibinfo {author}
  {\bibfnamefont {P.}~\bibnamefont {Bourges}},\ }\href
  {https://www.nature.com/articles/ncomms15119} {\bibfield  {journal} {\bibinfo
   {journal} {Nature communications}\ }\textbf {\bibinfo {volume} {8}},\
  \bibinfo {pages} {1} (\bibinfo {year} {2017})}\BibitemShut {NoStop}%
\bibitem [{\citenamefont {Murayama}\ \emph {et~al.}(2021)\citenamefont
  {Murayama}, \citenamefont {Ishida}, \citenamefont {Kurihara}, \citenamefont
  {Ono}, \citenamefont {Sato}, \citenamefont {Kasahara}, \citenamefont
  {Watanabe}, \citenamefont {Yanase}, \citenamefont {Cao}, \citenamefont
  {Mizukami}, \citenamefont {Shibauchi}, \citenamefont {Matsuda},\ and\
  \citenamefont {Kasahara}}]{PhysRevX.11.011021}%
  \BibitemOpen
  \bibfield  {author} {\bibinfo {author} {\bibfnamefont {H.}~\bibnamefont
  {Murayama}}, \bibinfo {author} {\bibfnamefont {K.}~\bibnamefont {Ishida}},
  \bibinfo {author} {\bibfnamefont {R.}~\bibnamefont {Kurihara}}, \bibinfo
  {author} {\bibfnamefont {T.}~\bibnamefont {Ono}}, \bibinfo {author}
  {\bibfnamefont {Y.}~\bibnamefont {Sato}}, \bibinfo {author} {\bibfnamefont
  {Y.}~\bibnamefont {Kasahara}}, \bibinfo {author} {\bibfnamefont
  {H.}~\bibnamefont {Watanabe}}, \bibinfo {author} {\bibfnamefont
  {Y.}~\bibnamefont {Yanase}}, \bibinfo {author} {\bibfnamefont
  {G.}~\bibnamefont {Cao}}, \bibinfo {author} {\bibfnamefont {Y.}~\bibnamefont
  {Mizukami}}, \bibinfo {author} {\bibfnamefont {T.}~\bibnamefont {Shibauchi}},
  \bibinfo {author} {\bibfnamefont {Y.}~\bibnamefont {Matsuda}}, \ and\
  \bibinfo {author} {\bibfnamefont {S.}~\bibnamefont {Kasahara}},\ }\href
  {\doibase 10.1103/PhysRevX.11.011021} {\bibfield  {journal} {\bibinfo
  {journal} {Phys. Rev. X}\ }\textbf {\bibinfo {volume} {11}},\ \bibinfo
  {pages} {011021} (\bibinfo {year} {2021})}\BibitemShut {NoStop}%
\bibitem [{\citenamefont {Liu}\ and\ \citenamefont
  {Dai}(2021{\natexlab{a}})}]{liu2021orbital}%
  \BibitemOpen
  \bibfield  {author} {\bibinfo {author} {\bibfnamefont {J.}~\bibnamefont
  {Liu}}\ and\ \bibinfo {author} {\bibfnamefont {X.}~\bibnamefont {Dai}},\
  }\href {https://www.nature.com/articles/s42254-021-00297-3#citeas} {\bibfield
   {journal} {\bibinfo  {journal} {Nature Reviews Physics}\ }\textbf {\bibinfo
  {volume} {3}},\ \bibinfo {pages} {367} (\bibinfo {year}
  {2021}{\natexlab{a}})}\BibitemShut {NoStop}%
\bibitem [{\citenamefont {Liu}\ \emph {et~al.}(2019)\citenamefont {Liu},
  \citenamefont {Ma}, \citenamefont {Gao},\ and\ \citenamefont
  {Dai}}]{PhysRevX.9.031021}%
  \BibitemOpen
  \bibfield  {author} {\bibinfo {author} {\bibfnamefont {J.}~\bibnamefont
  {Liu}}, \bibinfo {author} {\bibfnamefont {Z.}~\bibnamefont {Ma}}, \bibinfo
  {author} {\bibfnamefont {J.}~\bibnamefont {Gao}}, \ and\ \bibinfo {author}
  {\bibfnamefont {X.}~\bibnamefont {Dai}},\ }\href {\doibase
  10.1103/PhysRevX.9.031021} {\bibfield  {journal} {\bibinfo  {journal} {Phys.
  Rev. X}\ }\textbf {\bibinfo {volume} {9}},\ \bibinfo {pages} {031021}
  (\bibinfo {year} {2019})}\BibitemShut {NoStop}%
\bibitem [{\citenamefont {Liu}\ and\ \citenamefont
  {Dai}(2021{\natexlab{b}})}]{PhysRevB.103.035427}%
  \BibitemOpen
  \bibfield  {author} {\bibinfo {author} {\bibfnamefont {J.}~\bibnamefont
  {Liu}}\ and\ \bibinfo {author} {\bibfnamefont {X.}~\bibnamefont {Dai}},\
  }\href {\doibase 10.1103/PhysRevB.103.035427} {\bibfield  {journal} {\bibinfo
   {journal} {Phys. Rev. B}\ }\textbf {\bibinfo {volume} {103}},\ \bibinfo
  {pages} {035427} (\bibinfo {year} {2021}{\natexlab{b}})}\BibitemShut
  {NoStop}%
\bibitem [{\citenamefont {Bultinck}\ \emph {et~al.}(2020)\citenamefont
  {Bultinck}, \citenamefont {Khalaf}, \citenamefont {Liu}, \citenamefont
  {Chatterjee}, \citenamefont {Vishwanath},\ and\ \citenamefont
  {Zaletel}}]{PhysRevX.10.031034}%
  \BibitemOpen
  \bibfield  {author} {\bibinfo {author} {\bibfnamefont {N.}~\bibnamefont
  {Bultinck}}, \bibinfo {author} {\bibfnamefont {E.}~\bibnamefont {Khalaf}},
  \bibinfo {author} {\bibfnamefont {S.}~\bibnamefont {Liu}}, \bibinfo {author}
  {\bibfnamefont {S.}~\bibnamefont {Chatterjee}}, \bibinfo {author}
  {\bibfnamefont {A.}~\bibnamefont {Vishwanath}}, \ and\ \bibinfo {author}
  {\bibfnamefont {M.~P.}\ \bibnamefont {Zaletel}},\ }\href {\doibase
  10.1103/PhysRevX.10.031034} {\bibfield  {journal} {\bibinfo  {journal} {Phys.
  Rev. X}\ }\textbf {\bibinfo {volume} {10}},\ \bibinfo {pages} {031034}
  (\bibinfo {year} {2020})}\BibitemShut {NoStop}%
\bibitem [{\citenamefont {He}\ \emph {et~al.}(2012)\citenamefont {He},
  \citenamefont {Moore},\ and\ \citenamefont {Varma}}]{PhysRevB.85.155106}%
  \BibitemOpen
  \bibfield  {author} {\bibinfo {author} {\bibfnamefont {Y.}~\bibnamefont
  {He}}, \bibinfo {author} {\bibfnamefont {J.}~\bibnamefont {Moore}}, \ and\
  \bibinfo {author} {\bibfnamefont {C.~M.}\ \bibnamefont {Varma}},\ }\href
  {\doibase 10.1103/PhysRevB.85.155106} {\bibfield  {journal} {\bibinfo
  {journal} {Phys. Rev. B}\ }\textbf {\bibinfo {volume} {85}},\ \bibinfo
  {pages} {155106} (\bibinfo {year} {2012})}\BibitemShut {NoStop}%
\bibitem [{\citenamefont {Pozo}\ and\ \citenamefont
  {Souza}(2022)}]{pozo2022multipole}%
  \BibitemOpen
  \bibfield  {author} {\bibinfo {author} {\bibfnamefont {{\'O}.}~\bibnamefont
  {Pozo}}\ and\ \bibinfo {author} {\bibfnamefont {I.}~\bibnamefont {Souza}},\
  }\href {https://arxiv.org/abs/2211.10183} {\bibfield  {journal} {\bibinfo
  {journal} {arXiv preprint arXiv:2211.10183}\ } (\bibinfo {year}
  {2022})}\BibitemShut {NoStop}%
\end{thebibliography}%

\appendix

\begin{widetext}
\section{Details of the derivation of the intrinsic orbital ME tensor} \label{derivation_JJ}

The symmetric part $\Phi^{(\mathrm{S})ab}_{JJ}(\q,\omega)$ can be written as
\begin{eqnarray}
\Phi^{(\mathrm{S})ab}_{JJ} (\q,\omega)
=
 -e^2 \sum_{mn \kvec} \frac{f(\epsilon_{n\kvec+\q/2}) - f(\epsilon_{m\kvec-\q/2})}{\epsilon_{n\kvec+\q/2} - \epsilon_{m\kvec-\q/2} -( \omega + i\delta) } \R \Bigl( \bra{u_{m\kvec-\q/2}} v^a_{\kvec} \ket{u_{n\kvec+\q/2}} \bra{u_{n\kvec+\q/2}} v^b_{\kvec} \ket{u_{m\kvec-\q/2}} \Bigr).
\end{eqnarray}
Here, we neglect the diamagnetic term because it does not depend on $q$.
Using the following identity 
\begin{eqnarray}
\frac{1}{\Delta - \omega} = \frac{1}{\Delta} \Bigl( 1 + \frac{\omega}{\Delta - \omega}  \Bigr),
\end{eqnarray}
we obtain 
\begin{eqnarray}
\Phi^{(\mathrm{S})ab}_{JJ}(\q,\omega)
&=&
-e^2\sum_{mn \kvec} \frac{f(\epsilon_{n\kvec+\q/2}) - f(\epsilon_{m\kvec-\q/2})}{\epsilon_{n\kvec+\q/2} - \epsilon_{m\kvec-\q/2} - i\delta}
\biggl(
1 + \frac{\omega}{\epsilon_{n\kvec+\q/2} - \epsilon_{m\kvec-\q/2} -( \omega + i\delta)}
\biggr) \nonumber \\
&& \times \R \Bigl( \bra{u_{m\kvec-\q/2}} v^a_{\kvec} \ket{u_{n\kvec+\q/2}} \bra{u_{n\kvec+\q/2}} v^b_{\kvec} \ket{u_{m\kvec-\q/2}} \Bigr) . \label{Phi_symmetric}
\end{eqnarray}

In the following, we will expand the current-current correlation function up to the first order in $q$. For this purpose, we distinguish two cases; (i) the intraband case ($n=m$) and (ii) the interband case ($n \neq m$).

In the case of (i), the intraband case, we first derive the imaginary part of $\Phi^{(\mathrm{S})ab}_{JJ}(\q,\omega)$ and calculate the real part using the Kramers–Kronig relation. The imaginary part of the first term in  Eq.~(\ref{Phi_symmetric}) is proportional to
\begin{eqnarray}
(f(\epsilon_{n\kvec+\q/2}) - f(\epsilon_{m\kvec-\q/2})) \delta(\epsilon_{n\kvec+\q/2} - \epsilon_{m\kvec-\q/2}) = 0.
\end{eqnarray}
Thus, this term vanishes, and we need to consider only the second term. This term becomes
\begin{eqnarray}
\I \Phi^{(\mathrm{S-i})ab}_{JJ}(\q,\omega)
&\simeq&
-e^2 \sum_{n\kvec}  f'_n  \I \biggl( \frac{-\omega}{\omega + i\delta} - \frac{\omega (\partial_c \epsilon_{n\kvec}) q_c}{(\omega + i\delta)^2} \biggr)  (\partial_a \epsilon_{n\kvec}) (\partial_b \epsilon_{n\kvec}) \nonumber \\
&\simeq& \mathcal{O}(q^0)
-e^2 \sum_{n\kvec} \I \biggl(
\frac{ - f'_n \omega}{(\omega + i\delta)^2} \biggr) (\partial_c \epsilon_{n\kvec}) (\partial_a \epsilon_{n\kvec}) (\partial_b \epsilon_{n\kvec}) q_c  ,
\end{eqnarray}
where $f'_n = \partial f(\epsilon_{n\kvec})/\partial \epsilon_{n\kvec}$.
Using the Kramers-Kronig relation, we obtain
\begin{eqnarray}
\Phi^{(\mathrm{S-i})ab}_{JJ}(\q,\omega)
&\simeq& \mathcal{O}(q^0)
- e^2 \sum_{n\kvec} 
\frac{ - f'_n \omega}{(\omega + i\delta)^2}  (\partial_c \epsilon_{n\kvec}) (\partial_a \epsilon_{n\kvec}) (\partial_b \epsilon_{n\kvec}) q_c .
\end{eqnarray}
The intraband term is totally symmetric for $a$, $b$, and $c$ interchanges. Thus, this term only contributes to the electric quadrupole response. 

Next, we consider the interband case (ii). In this work, we focus on the ME effect in the static limit ($ \lim_{\omega \to 0} \beta_{ij}/i\omega $). Thus, we need to calculate $\lim_{\omega \to 0} \R \Phi^{ab,c}_{JJ}(0,\omega)/\omega$. We will calculate this quantity directly in the interband case (ii). 
In the interband case, we can take $\omega, \delta \to 0$ first because there is an energy gap in the denominator, and singularities do not exist. At first, we will calculate the contribution from the first term in the Eq.~(\ref{Phi_symmetric}), which is proportional to
\begin{eqnarray}
\sum_{m\neq n \kvec} \frac{f(\epsilon_{n\kvec+\q/2}) - f(\epsilon_{m\kvec-\q/2})}{\epsilon_{n\kvec+\q/2} - \epsilon_{m\kvec-\q/2}} 
\R \Bigl( \bra{u_{m\kvec-\q/2}} v^a_{\kvec} \ket{u_{n\kvec+\q/2}} \bra{u_{n\kvec+\q/2}} v^b_{\kvec} \ket{u_{m\kvec-\q/2}} \Bigr).
\end{eqnarray}
This equation is even for $\q \to -\q$. The first-order term in $q$ becomes zero after summing over $n$ and $m$. Thus, we need to consider only the first order in $q$ contribution from the second term. The second term is 
\begin{eqnarray}
\lim_{\omega \to 0} \R \Phi^{(\mathrm{S-ii})ab}(\q,\omega)/\omega 
=
-e^2\sum_{m \neq n \kvec}
\frac{f(\epsilon_{n\kvec+\q/2}) - f(\epsilon_{m\kvec-\q/2})}{(\epsilon_{n\kvec+\q/2} - \epsilon_{m\kvec-\q/2})^2} 
\R \Bigl( \bra{u_{m\kvec-\q/2}} v^a_{\kvec} \ket{u_{n\kvec+\q/2}} \bra{u_{n\kvec+\q/2}} v^b_{\kvec} \ket{u_{m\kvec-\q/2}} \Bigr) . \nonumber \\
\end{eqnarray}
Expanding each coefficients in $q$, we obtain
\begin{eqnarray}
&&\frac{f(\epsilon_{n\kvec+\q/2}) - f(\epsilon_{m\kvec-\q/2})}{(\epsilon_{n\kvec+\q/2} - \epsilon_{m\kvec-\q/2})^2}
\simeq
\frac{f_{nm}}{(\epsilon_{nm\kvec})^2} + q_c \biggl(
\frac{\partial_c \tilde{f}_{nm}}{2 (\epsilon_{nm\kvec})^2}
-
\frac{f_{nm} (\partial_c \tilde{\epsilon}_{nm\kvec})}{(\epsilon_{nm\kvec})^3}
\biggr) \label{expand1} \\
&&\R \Bigl( \bra{u_{m\kvec-\q/2}} v^a_{\kvec} \ket{u_{n\kvec+\q/2}} \bra{u_{n\kvec+\q/2}} v^b_{\kvec} \ket{u_{m\kvec-\q/2}} \Bigr) \nonumber \\
&&\simeq
(\epsilon_{nm\kvec})^2 \R [\A^a_{mn} \A^b_{nm}] 
-
\frac{q_c}{2} \epsilon_{nm\kvec} \R [( \bra{u_{m\kvec}} v^a_{\kvec} \ket{\partial_c u_{n\kvec}} - \bra{\partial_c u_{m\kvec}} v^a_{\kvec} \ket{u_{n\kvec}} ) \braket{u_{n\kvec} | \partial_b u_{m\kvec}} + (a \leftrightarrow b)  ], \label{expand2}
\end{eqnarray}
where $f_{nm} = f(\epsilon_{n\kvec}) - f(\epsilon_{m\kvec})$, $\tilde{f}_{nm} = f(\epsilon_{n\kvec}) + f(\epsilon_{m\kvec})$, $\epsilon_{nm\kvec}=\epsilon_{n\kvec}-\epsilon_{m\kvec}$, and $\tilde{\epsilon}_{nm\kvec}=\epsilon_{n\kvec}+\epsilon_{m\kvec}$.
$\A^{a}_{mn} \equiv i \braket{ u_{m\kvec}  | \partial_a u_{n\kvec} }$ is the Berry connection.  
Considering the first-order terms in $q$ in Eq.~(\ref{expand1}), we obtain (a) (In the following, we neglect the 0th order of $q$)
\begin{eqnarray}
\lim_{\omega \to 0} \R \Phi^{(\mathrm{S-ii-a})ab}(\q,\omega)/\omega 
&=&
- e^2\sum_{m \neq n \kvec} \biggl( \frac{\partial_c \tilde{f}_{nm}}{2} \R [\A^a_{mn} \A^b_{nm}] 
-
\frac{f_{nm}}{\epsilon_{nm\kvec}} (\partial_c \tilde{\epsilon}_{nm\kvec}) \R [\A^a_{mn} \A^b_{nm}] \biggr) q_c \nonumber \\
&=&
- e^2\sum_{n \kvec} \sum_{m(\neq n)} \biggl(
\partial_c f_n \R[\A^a_{mn} \A^b_{nm}]
-
\frac{2f_n}{\epsilon_{nm\kvec}}(\partial_c \tilde{\epsilon}_{nm\kvec}) \R[\A^a_{mn} \A^b_{nm}]
\biggr) q_c .
\end{eqnarray}

Next, considering the first order term in $q$ in Eq.~(\ref{expand2}), we obtain (b)
\begin{eqnarray}
\lim_{\omega \to 0} \R \Phi^{(\mathrm{S-ii-b})ab}_{JJ}(\q, \omega)/\omega 
&=&
 \frac{e^2}{2} \sum_{m\neq n \kvec} \frac{f_{nm}}{\epsilon_{nm\kvec}}
\R \Bigl[\Bigl( \bra{u_{m\kvec}} v^a_{\kvec} \ket{\partial_c u_{n\kvec}} - \bra{\partial_c u_{m\kvec}} v^a_{\kvec} \ket{u_{n\kvec}} \Bigr) \braket{u_{n\kvec} | \partial_b u_{m\kvec}} + (a \leftrightarrow b)  \Bigr] q_c\nonumber \\
&=& 
e^2 \sum_{m \neq n \kvec} \frac{f_n}{\epsilon_{nm\kvec}}
\R \Bigl[\Bigl( \bra{u_{m\kvec}} v^a_{\kvec} \ket{\partial_c u_{n\kvec}} - \bra{\partial_c u_{m\kvec}} v^a_{\kvec} \ket{u_{n\kvec}} \Bigr) \braket{u_{n\kvec} | \partial_b u_{m\kvec}} + (a \leftrightarrow b)  \Bigr] q_c . \nonumber \\ \label{JJ_ii_b}
\end{eqnarray}
If we focus on the wave function part in the Eq.~(\ref{JJ_ii_b}), we can write
\begin{eqnarray}
R_{abc}
&=&\R \Bigl[\Bigl( \bra{u_{m\kvec}} v^a_{\kvec} \ket{\partial_c u_{n\kvec}} - \bra{\partial_c u_{m\kvec}} v^a_{\kvec} \ket{u_{n\kvec}} \Bigr) \braket{u_{n\kvec} | \partial_b u_{m\kvec}} \Bigr] + (a \leftrightarrow b) \nonumber \\
&=&
\sum_l \R \Bigl[ (\bra{u_{m\kvec}} v^a_{\kvec} \ket{u_{l\kvec}} \braket{ u_{l\kvec} | \partial_c u_{n\kvec} } 
-
\braket{\partial_c u_{m\kvec}| u_{l\kvec}} \bra{u_{l\kvec}} v^a_{\kvec} \ket{u_{n\kvec}}) \braket{u_{n\kvec} | \partial_b u_{m\kvec}} \Bigr] + (a \leftrightarrow b) \nonumber \\
&=&\Bigl(
(\partial_a \tilde{\epsilon}_{nm\kvec}) \R [\braket{u_{m\kvec} | \partial_c u_{n\kvec}}
\braket{ u_{n\kvec} | \partial_b u_{m\kvec}}
] \nonumber \\
&&
+
\sum_{l(\neq m)} \epsilon_{lm\kvec} \R[ \braket{u_{m\kvec}  | \partial_a u_{l\kvec}  } \braket{ u_{l\kvec} | \partial_c u_{n\kvec}} \braket{u_{n\kvec} | \partial_b u_{m\kvec} } ] \nonumber \\
&&
-
\sum_{l(\neq n)} \epsilon_{nl\kvec} \R [\braket{ \partial_c u_{m\kvec} | u_{l\kvec}} \braket{ u_{l\kvec} | \partial_a u_{n\kvec} } \braket{u_{n\kvec} | \partial_b u_{m\kvec}}  ]
\Bigr) + (a \leftrightarrow b) .
\end{eqnarray}
We write $\epsilon_{nl \kvec}$ as $\epsilon_{nl \kvec} = \epsilon_{nm \kvec} + \epsilon_{ml\kvec}$ and calculate $R_{abc}$ for the case containing $\epsilon_{nm\kvec}$ (p), the case containing $\epsilon_{lm\kvec}$ (q), and  the case neither containing $\epsilon_{nm\kvec}$ or $\epsilon_{lm\kvec}$ (r). In the case of (p), we obtain
\begin{eqnarray}
R^{(\mathrm{p})}_{abc}
&=& \Bigl(
\epsilon_{nm\kvec} \R [\braket{u_{m\kvec} | \partial_a u_{n\kvec} }  \braket{ u_{n\kvec} | \partial_c u_{n\kvec} } \braket{ u_{n\kvec} | \partial_b u_{m\kvec}} ] \nonumber \\
&&
-\epsilon_{nm\kvec}
\sum_{l(\neq n)} 
\R [\braket{ \partial_c u_{m\kvec} | u_{l\kvec}} \braket{ u_{l\kvec} | \partial_a u_{n\kvec} } \braket{u_{n\kvec} | \partial_b u_{m\kvec}}  ]
\Bigr) + (a \leftrightarrow b) \nonumber \\
&=&\epsilon_{nm\kvec} \Bigl(
\R [\braket{u_{m\kvec} | \partial_a u_{n\kvec} }  \braket{ u_{n\kvec} | \partial_c u_{n\kvec} } \braket{ u_{n\kvec} | \partial_b u_{m\kvec}} ] \nonumber \\
&&-
\R [\braket{u_{m\kvec} | \partial_c u_{n\kvec} }  \braket{ u_{n\kvec} | \partial_a u_{n\kvec} } \braket{ u_{n\kvec} | \partial_b u_{m\kvec}} ] \nonumber \\
&& -
\R [ \braket{ \partial_c u_{m\kvec} | \partial_a u_{n\kvec} }
\braket{u_{n\kvec} | \partial_b u_{m\kvec}}
]
\Bigr) + (a \leftrightarrow b).
\end{eqnarray}
This term contributes to $\Phi_{JJ}$ as
\begin{eqnarray}
\lim_{\omega \to 0} \R \Phi^{(\mathrm{S-ii-b-p})ab}_{JJ}(\q, \omega)/\omega 
&=&
e^2 \sum_{n\kvec} \sum_{m(\neq n)} \frac{f_n}{\epsilon_{nm\kvec}} R^{(\mathrm{p})}_{abc} q_c \nonumber \\
&=& e^2 \sum_{n\kvec} f_n
\R \Bigl(
-\braket{\partial_b u_{n\kvec} | \partial_a u_{n\kvec}}
\braket{u_{n\kvec} | \partial_c u_{n\kvec}}
+
\braket{ \partial_b u_{n\kvec} | u_{n\kvec}} \braket{u_{n\kvec} | \partial_a u_{n\kvec}} \braket{u_{n\kvec} | \partial_c u_{n\kvec}} \nonumber \\
&&
+ \braket{\partial_b u_{n\kvec} | \partial_c u_{n\kvec}}
\braket{u_{n\kvec} | \partial_a u_{n\kvec}}
-
\braket{\partial_b u_{n\kvec} | u_{n\kvec}}
\braket{u_{n\kvec} | \partial_c u_{n\kvec}}
\braket{u_{n\kvec} | \partial_a u_{n\kvec}} \nonumber \\
&& +
\sum_{m(\neq n)}
\braket{\partial_c u_{m\kvec} | \partial_a u_{n\kvec}}
\braket{\partial_b u_{n\kvec} |  u_{m\kvec}}
\Bigr) q_c + (a \leftrightarrow b) \nonumber \\
&=&
e^2 \sum_{n\kvec } f_n \R \Bigl( - \bra{\partial_b u_{n\kvec}} \partial_c Q_n \ket{\partial_a u_{n\kvec}}
+
\bra{\partial_b u_{n\kvec}} \partial_c Q_n \ket{\partial_a u_{n\kvec}}
\Bigr) q_c \nonumber \\
&=& 0.
\end{eqnarray}
Thus, $R^{(\mathrm{p})}_{abc}$  vanishes. Here, we have defined $Q_n = 1- \ket{u_{n\kvec}}\bra{u_{n\kvec}}$.

In the case of (q),
\begin{eqnarray}
R^{(\mathrm{q})}_{abc}
&=&
\sum_{l(\neq n)} \epsilon_{lm\kvec} \R \Bigl(
\braket{u_{m\kvec} | \partial_a u_{l\kvec}}
\braket{u_{l\kvec} | \partial_c u_{n\kvec}}
\braket{u_{n\kvec} | \partial_b u_{m\kvec}}
-
\braket{u_{m\kvec} | \partial_c u_{l\kvec}}
\braket{u_{l\kvec} | \partial_a u_{n\kvec}}
\braket{u_{n\kvec} | \partial_b u_{m\kvec}}
\Bigr) \nonumber \\
&& + (a \leftrightarrow b) \nonumber \\
&=&
\sum_{l(\neq n)} \R \Bigl(
-\tilde{V}^a_{ml} \A^c_{ln} \A^b_{nm}
+\tilde{V}^c_{ml} \A^a_{ln} \A^b_{nm}
\Bigr) + (a \leftrightarrow b),
\end{eqnarray}
where $\tilde{V}^a_{ml} \equiv v^a_{ml} - v^{0a}_m \delta_{ml} $. Adding $R^{(\mathrm{r})}_{abc}$ to $R^{\mathrm{(q)}}_{abc}$, we find
\begin{eqnarray}
R^{\mathrm{(q)}}_{abc} + R^{\mathrm{(r)}}_{abc}
&=&
\sum_{l(\neq n)} \R \Bigl(
- V^a_{ml,n} \A^c_{ln} \A^b_{nm}
+ \tilde{V}^c_{ml} \A^a_{ln} \A^b_{nm}
\Bigr) + (a \leftrightarrow b),
\end{eqnarray}
where $V^a_{ml,n} \equiv v^a_{ml} + v^{0a}_n \delta_{ml}  $. Finally, by combining (q) and (r), we get
\begin{eqnarray}
&&\lim_{\omega \to 0} \R \Phi^{(\mathrm{S-ii-b-q})ab}_{JJ}(\q, \omega)/\omega
+
\lim_{\omega \to 0} \R \Phi^{(\mathrm{S-ii-b-r})ab}_{JJ}(\q, \omega)/\omega  \nonumber \\
&&=
e^2 \sum_{n \kvec} \sum_{m(\neq n)} \frac{f_n}{\epsilon_{nm\kvec}}
\sum_{l(\neq n)} \Bigl[ \R \Bigl(
- V^a_{ml,n} \A^c_{ln} \A^b_{nm}
+ \tilde{V}^c_{ml} \A^a_{ln} \A^b_{nm}
\Bigr) + (a \leftrightarrow b) \Bigr]q_c.
\end{eqnarray}
Adding (a) to this equation, we obtain
\begin{eqnarray}
&&\lim_{\omega \to 0} \R \Phi^{(\mathrm{S-ii-b-q})ab}_{JJ}(\q, \omega)/\omega
+
\lim_{\omega \to 0} \R \Phi^{(\mathrm{S-ii-b-r})ab}_{JJ}(\q, \omega)/\omega
+
\lim_{\omega \to 0} \R \Phi^{(\mathrm{S-ii-a})ab}_{JJ}(\q, \omega)/\omega
\nonumber \\
&&= e^2\sum_{n\kvec} f_n \biggl(
\partial_c g^{ab}_n
+ 
\sum_{m(\neq n)} \sum_{l(\neq n)}
\frac{1}{\epsilon_{nm\kvec}}
\R \Bigl(
-V^a_{ml,n} \A^c_{ln} \A^b_{nm}
-V^b_{ml,n} \A^c_{ln} \A^a_{nm}
+V^c_{ml,n} \A^a_{ln} \A^b_{nm}
+V^c_{ml,n} \A^b_{ln} \A^a_{nm}
\Bigr)
\biggr) q_c  \nonumber \\
&& \equiv \lim_{\omega \to 0} \Phi^{(\mathrm{S})ab,c}_{JJ} q_c/\omega ,
\end{eqnarray}
where $g^{ab}_n = \sum_{m(\neq n)} \R [ \A^a_{nm} \A^b_{mn}]$ is the quantum metric.

Finally, we calculate the orbital ME tensor ${\chi}^{(\mathrm{me})}_{da} \equiv - \varepsilon_{bca} \lim_{\omega \to 0} \Phi^{(\mathrm{S})db,c}_{JJ}/3\omega $ as
\begin{eqnarray}
\chi^{(\mathrm{me})}_{da} 
= 
-e^2 \sum_{n\kvec} f_n \biggl(
\frac{1}{3} \varepsilon_{bca} \partial_{c} g^{db}_n - \sum_{m(\neq n)} \frac{2}{\epsilon_{nm\kvec}}
\R \Bigl[ 
\A^d_{nm} M^a_{mn}
-
\frac{1}{3} \delta_{da} \A^b_{nm} M^b_{mn} 
\Bigr]
\biggr) \label{mainresult},
\end{eqnarray}
where $\bm{M}_{mn} = \sum_{l(\neq n)} \bm{V}_{ml,n}/2  \times \bm{\A}_{ln}$ behaves as a off-diagonal orbital magnetization. 

\section{Derivation of the intrinsic orbital ME effect based on a scalar potential} \label{derivation_Jn}
In the main text, we have derived the orbital ME tensor using the current-current correlation function. In this Appendix, we will show that the current-density correlation function also contains information on the ME effect.

We introduce a scalar potential $\phi(\x,t) = \phi_{\q,\omega} e^{-i\omega t + i\q \cdot \x}$ as an external field
\begin{eqnarray}
H_{\phi} = -e\phi(\x,t) n(\x).
\end{eqnarray}
The scalar potential induces an electric field $\bm{E}_{\q,\omega} = -i\q \phi_{\q,\omega}$. A current changes linearly when applying an electric field as
\begin{eqnarray}
J^i_{\q,\omega} = \Phi^i_{Jn}(\q,\omega) \phi_{\q,\omega} \label{jn_Kubo}.
\end{eqnarray}
Expanding the current-density correlation function $\Phi^i_{Jn}(\q,\omega)$ by $q$ up to the second order, we focus on the second derivative of the function $\Phi^{i,jk}_{Jn}(\omega) = \partial_{q_j q_k} \Phi^i_{Jn}(0,\omega)$. This function is symmetric for the interchange $j \leftrightarrow k$, so we can decompose this function using a traceless rank-2 tensor $\tilde{\beta}_{ij}$ and a totally symmetric rank-3 tensor $\tilde{\gamma}_{ijk}$ as
\begin{subequations}
\begin{align}
& \Phi^{k,ij}_{Jn}(\omega)
=
i \varepsilon_{jkl} \tilde{\beta}_{il}(\omega) + i \varepsilon_{ikl} \tilde{\beta}_{jl}(\omega) + \omega \tilde{\gamma}_{ijk}(\omega) \label{jn_correlation} \\
& \tilde{\beta}_{li}(\omega) = \frac{1}{3i} \varepsilon_{ijk} \Phi^{k,lj}_{Jn}(\omega) \label{Jn_me_response} \\
& \tilde{\gamma}_{ijk}(\omega) = \frac{1}{3\omega} \Bigl( \Phi^{i,jk}_{Jn}(\omega)+\Phi^{j,ki}_{Jn}(\omega)+\Phi^{k,ij}_{Jn}(\omega) \Bigr) .
\end{align}
\end{subequations}
Substituting Eq.~(\ref{jn_correlation}) into Eq.~(\ref{jn_Kubo}), we find
\begin{eqnarray}
J^k_{\q,\omega} &=& \Phi^{k,ij}_{Jn} q_i q_j \phi_{\q,\omega} \nonumber \\
&=&
\Bigl( i \q \times (-2i\tilde{\beta}^{\mathrm{t}}(\omega) \bm{E}_{\q,\omega}) \Bigr)_k + i q_i \omega  \tilde{\gamma}_{ijk}(\omega) E^k_{\q,\omega}.
\end{eqnarray}
We need to calculate  $\tilde{\beta}_{ij}$ to obtain the orbital ME tensor.

According to the linear response theory, the current-density correlation function is 
\begin{eqnarray}
\Phi^{i}_{Jn}(\q,\omega)
=
e^2 \sum_{mn,\kvec} \frac{f(\epsilon_{n\kvec+\q/2}) - f(\epsilon_{m\kvec-\q/2})}{\epsilon_{n\kvec+\q/2} - \epsilon_{m\kvec-\q/2} -( \omega +  i\delta )} \times 
\bra{u_{m\kvec-\q/2}} v^i_{\kvec} \ket{u_{n\kvec+\q/2}} \braket{u_{n\kvec+\q/2} | u_{m\kvec-\q/2}} .
\end{eqnarray}
In the following, we evaluate this function dividing it into two cases; (A) an intraband case ($m=n$) and (B) an interband case ($m\neq n$).

In the case of (A), the correlation function becomes
\begin{eqnarray}
\Phi^{i(\mathrm{A})}_{Jn}(\q,\omega)
=
e^2\sum_{n,\kvec} \frac{f(\epsilon_{n\kvec+\q/2}) - f(\epsilon_{n\kvec-\q/2})}{\epsilon_{n\kvec+\q/2} - \epsilon_{n\kvec-\q/2} -( \omega + i\delta)} \times 
\bra{u_{n\kvec-\q/2}} v^i_{\kvec} \ket{u_{n\kvec+\q/2}} \braket{u_{n\kvec+\q/2} | u_{n\kvec-\q/2}}. 
\end{eqnarray}
Expanding each coefficient by $q$ up to the second order, we obtain
\begin{eqnarray}
&&\frac{f(\epsilon_{n\kvec+\q/2}) - f(\epsilon_{n\kvec-\q/2})}{\epsilon_{n\kvec+\q/2} - \epsilon_{n\kvec-\q/2} - (\omega + i\delta)}
\simeq 
\frac{-f'_n (\partial_a \epsilon_n) q_a}{\omega + i\delta} - \frac{f'_n (\partial_a \epsilon_{n}) (\partial_b \epsilon_n) q_a q_b}{(\omega + i\delta)^2}
\\
&&\braket{u_{n\kvec+\q/2} | u_{n\kvec-\q/2}}
\simeq
1 - q_a \braket{u_{n\kvec} | \partial_a u_{n\kvec}} \\
&&\bra{u_{n\kvec-\q/2}} v^i_{\kvec}  \ket{u_{n\kvec+\q/2}}
\simeq
(\partial_i \epsilon_{n})
+
\frac{q_a}{2}( \bra{u_{n\kvec}} v^i_{\kvec} \ket{\partial_a u_{n\kvec}}
-
\bra{\partial_a u_{n\kvec}} v^i_{\kvec} \ket{ u_{n\kvec}}
).
\end{eqnarray}
Thus, the second derivative of $\Phi^{i}_{Jn}(\q,\omega)$ is
\begin{eqnarray}
\Phi^{i,ab(\mathrm{A})}_{Jn}(\omega) q_aq_b
&&=
q_a q_b e^2 \sum_{n,\kvec} \biggl[\frac{-f'_n (\partial_a \epsilon_n) }{\omega +i\delta}  \Bigl(  - (\partial_i \epsilon_n) \braket{u_{n\kvec} | \partial_b u_{n\kvec}} 
+
\frac{1}{2} ( \bra{u_{n\kvec}} v^i_{\kvec} \ket{\partial_b u_{n\kvec}}
-
\bra{\partial_b u_{n\kvec}} v^i_{\kvec} \ket{ u_{n\kvec}})
\Bigr) \nonumber \\
&&\hspace{1.8cm}
-\frac{f'_n}{( \omega +i\delta)^2}
(\partial_a \epsilon_n) (\partial_b \epsilon_n) (\partial_i \epsilon_n) \biggr] .
\end{eqnarray}
The first term is proportional to $1/(\omega + i\delta)^1$ and can be transformed to
\begin{eqnarray}
- (\partial_i \epsilon_n) \braket{u_{n\kvec} | \partial_b u_{n\kvec}} 
+
\frac{1}{2} ( \bra{u_{n\kvec}} v^i_{\kvec} \ket{\partial_b u_{n\kvec}}
-
\bra{\partial_b u_{n\kvec}} v^i_{\kvec} \ket{ u_{n\kvec}})
&=&
\frac{1}{2} \Bigl(  
\bra{u_{n\kvec}} v^i_{\kvec} Q_n \ket{\partial_b u_{n\kvec}}
-
\bra{\partial_b u_{n\kvec}} Q_n v^i_{\kvec} \ket{u_{n\kvec}}
\Bigr) \nonumber \\
&=&
\frac{1}{2} \Bigl(
\bra{\partial_i u_{n\kvec}} \epsilon_n -H_{\kvec} \ket{\partial_b u_{n\kvec} } - \mathrm{c.c.}
\Bigr). \nonumber \\
&\equiv& m_n^{ib} - (\mathrm{c.c.}).
\end{eqnarray}
Using this relationship, we obtain
\begin{eqnarray}
\Phi^{i,ab(\mathrm{A})}_{Jn}(\omega) q_aq_b 
&=&
q_a q_b e^2\sum_{n,\kvec} \biggl[
\frac{-f'_n}{\omega +i\delta} (\partial_a \epsilon_n) \bigl( m^{ib}_n - m^{bi}_n \bigr)
-
\frac{f'_n}{(\omega + i\delta)^2} (\partial_a \epsilon_n) (\partial_b \epsilon_n) (\partial_i \epsilon_n)
\biggr].
\end{eqnarray}

Next, we calculate the case of (B). In this case, the current-density correlation function becomes
\begin{eqnarray}
\Phi^{i(\mathrm{B})}_{Jn}(\q,\omega)
&=&
e^2 \sum_{m \neq n ,\kvec} \frac{f(\epsilon_{n\kvec+\q/2}) - f(\epsilon_{m\kvec-\q/2})}{\epsilon_{n\kvec+\q/2} - \epsilon_{m\kvec-\q/2} -( \omega +i\delta)}
\times
\bra{u_{m\kvec-\q/2}} v^i_{\kvec} \ket{u_{n\kvec+\q/2}} \braket{ u_{n\kvec+\q/2} | u_{m\kvec-\q/2}}.
\end{eqnarray}
In the interband case, the denominator does not have a singularity, so we take the limit $\omega \to 0$ and $i \delta \to 0$. Expanding each coefficient by $q$ up to the second order, we get
\begin{eqnarray}
\braket{u_{n\kvec+\q/2} | u_{m\kvec-\q/2}}
&\simeq&
-q_a \braket{u_{n\kvec} | \partial_a u_{m\kvec}}
-\frac{q_a q_b}{2} \braket{ \partial_a u_{n\kvec} | \partial_b u_{m\kvec}} \\
\bra{u_{m\kvec-\q/2}} v^i_{\kvec} \ket{u_{n\kvec+\q/2}}
&\simeq&
i \epsilon_{mn\kvec} \A^i_{mn}
-\frac{q_a}{2}( \bra{\partial_a u_{m\kvec}} v^i_{\kvec} \ket{u_{n\kvec}}
- \bra{u_{m\kvec}} v^i_{\kvec} \ket{\partial_a u_{n\kvec}}
) \nonumber \\
&=&
i \epsilon_{mn\kvec} \A^i_{mn}
-\frac{q_a}{2} \Bigl(
- \partial_i \tilde{\epsilon}_{nm\kvec} \braket{u_{m\kvec} | \partial_a u_{n\kvec}} \nonumber \\
&& + \sum_{l(\neq n)} \braket{\partial_a u_{m\kvec} | u_{l\kvec}} i \epsilon_{ln\kvec} \A^i_{ln}
- \sum_{l (\neq m)} i \epsilon_{ml\kvec} \A^i_{ml} \braket{u_{l\kvec} | \partial_a u_{n\kvec}}
\Bigr) \\
\frac{f(\epsilon_{n\kvec+\q/2}) - f(\epsilon_{m\kvec-\q/2})}{\epsilon_{n\kvec+\q/2} - \epsilon_{m\kvec-\q/2}}
&\simeq&
\frac{f_{nm}}{\epsilon_{nm\kvec}}
+
\frac{q_a}{2 \epsilon_{nm\kvec}} \biggl( 
\partial_a \tilde{f}_{nm} - \frac{(\partial_a \tilde{\epsilon}_{nm\kvec}) f_{nm}}{\epsilon_{nm\kvec}}
\biggr).
\end{eqnarray}
Collecting the second-order terms, we obtain
\begin{eqnarray}
\Phi^{i,ab(\mathrm{B})}_{Jn}(\omega=0) q_a q_b
&=&
e^2
\sum_{m \neq n \kvec} \biggl\{
-
\frac{f_{nm}}{2} \braket{u_{m\kvec} | \partial_i u_{n\kvec}} \braket{\partial_a u_{n\kvec} | \partial_b u_{m\kvec}} 
+
\frac{f_{nm}}{2 \epsilon_{nm\kvec}} \braket{u_{n\kvec} | \partial_a u_{m\kvec}}
\Bigl(
- \partial_i \tilde{\epsilon}_{nm\kvec} \braket{u_{m\kvec} | \partial_b u_{n\kvec}}
\nonumber \\
&&
-
\sum_{l(\neq n)} \braket{\partial_b u_{m\kvec} | u_{l\kvec}} \epsilon_{ln \kvec} \braket{u_{l\kvec} | \partial_i u_{n\kvec}}
+
\sum_{l(\neq m)} \epsilon_{ml\kvec} \braket{u_{m\kvec} | \partial_i u_{l\kvec}} \braket{ u_{l\kvec} | \partial_b u_{n\kvec}}
\Bigr) \nonumber \\
&&
-\frac{1}{2} \braket{u_{n\kvec} | \partial_a u_{m\kvec}} \braket{u_{m\kvec} | \partial_i u_{n\kvec}}
\biggl(
\partial_b \tilde{f}_{nm} - \frac{(\partial_b \tilde{\epsilon}_{nm\kvec}) f_{nm}}{\epsilon_{nm\kvec}}
\biggr)
\biggr\} q_a q_b \nonumber \\
&=&
e^2
\sum_{m \neq n \kvec} \biggl\{
-
f_n \R  [\braket{u_{m\kvec} | \partial_i u_{n\kvec}} \braket{\partial_a u_{n\kvec} | \partial_b u_{m\kvec}} ]
+
\frac{f_n}{ \epsilon_{nm\kvec}} 
\Bigl(
- \partial_i \tilde{\epsilon}_{nm\kvec} \R [ \braket{u_{n\kvec} | \partial_a u_{m\kvec}} \braket{u_{m\kvec} | \partial_b u_{n\kvec}} ]
\nonumber \\
&&
-
\sum_{l(\neq n)}  \epsilon_{ln\kvec} \R [\braket{u_{n\kvec} | \partial_a u_{m\kvec}} \braket{\partial_b u_{m\kvec} | u_{l\kvec}} \braket{u_{l\kvec} | \partial_i u_{n\kvec}} ] 
\nonumber \\
&&
+
\sum_{l(\neq m)}  \epsilon_{ml\kvec} \R [ \braket{u_{n\kvec} | \partial_a u_{m\kvec}} \braket{u_{m\kvec} | \partial_i u_{l\kvec}} \braket{ u_{l\kvec} | \partial_b u_{n\kvec}} ]
\Bigr) \nonumber \\
&&
- \R[ \braket{u_{n\kvec} | \partial_a u_{m\kvec}} \braket{u_{m\kvec} | \partial_i u_{n\kvec}} ]
\biggl(
\partial_b f_{n} - \frac{(\partial_b \tilde{\epsilon}_{nm\kvec}) f_{n}}{\epsilon_{nm\kvec}}
\biggr)
\biggr\} q_a q_b.
\end{eqnarray}
In the following, we separate the calculation into two cases; there is (i) $(\epsilon_{nm\kvec})^0$ or (ii) $(\epsilon_{nm\kvec})^1$ in the denominator.
In the case of (i)
\begin{eqnarray}
\Phi^{i,ab(\mathrm{B-i})}_{Jn}(0)q_aq_b
&=&
e^2
\sum_{m \neq n \kvec} f_n  \biggl\{
-
\R  [\braket{u_{m\kvec} | \partial_i u_{n\kvec}} \braket{\partial_a u_{n\kvec} | \partial_b u_{m\kvec}} ]
+
\sum_{l(\neq n)} \R [\braket{u_{n\kvec} | \partial_a u_{m\kvec}} \braket{\partial_b u_{m\kvec} | u_{l\kvec}} \braket{u_{l\kvec} | \partial_i u_{n\kvec}} ]  \nonumber \\
&&
-
\R [ \braket{u_{n\kvec} | \partial_a u_{m\kvec}} \braket{u_{m\kvec} | \partial_i u_{n\kvec}} \braket{ u_{n\kvec} | \partial_b u_{n\kvec}} ]
+
\partial_b \R[ \braket{u_{n\kvec} | \partial_a u_{m\kvec}} \braket{u_{m\kvec} | \partial_i u_{n\kvec}} ] \biggr\} q_a q_b .
\end{eqnarray}
The second term can be rewritten as 
\begin{eqnarray}
&&\text{the second term of }\Phi^{i,ab(\mathrm{B-i})}_{Jn}(0)q_aq_b \nonumber \\
&&=
e^2 \sum_{n \kvec} \sum_{m(\neq n),l(\neq n)} f_n \R [\braket{u_{n\kvec} | \partial_a u_{m\kvec}} \braket{\partial_b u_{m\kvec} | u_{l\kvec}} \braket{u_{l\kvec} | \partial_i u_{n\kvec}} ] q_a q_b \nonumber \\
&&=
e^2 \sum_{n \kvec} \sum_{m(\neq n),l(\neq n)} f_n \R [\braket{u_{n\kvec} | \partial_a u_{l\kvec}} \braket{\partial_b u_{l\kvec} | u_{m\kvec}} \braket{u_{m\kvec} | \partial_i u_{n\kvec}} ] q_a q_b \nonumber \\
&&=
e^2 \sum_{n \kvec} \sum_{m(\neq n)} f_n \Bigl(
\R [\braket{\partial_a  u_{n\kvec} |\partial_b  u_{m\kvec}} \braket{u_{m\kvec} | \partial_i u_{n\kvec}} ]
-
\R [\braket{u_{n\kvec} | \partial_a u_{n\kvec}} \braket{\partial_b u_{n\kvec} | u_{m\kvec}} \braket{u_{m\kvec} | \partial_i u_{n\kvec}} ]
\Bigr)q_a q_b .
\end{eqnarray}
This term cancels out with the first and third terms. Thus, the final remaining term is
\begin{eqnarray}
\Phi^{i,ab(\mathrm{B-i})}_{Jn}(0)q_aq_b
&=&
e^2 \sum_{m \neq n \kvec} f_n \partial_b \R[ \braket{u_{n\kvec} | \partial_a u_{m\kvec}} \braket{u_{m\kvec} | \partial_i u_{n\kvec}} ] q_a q_b \nonumber \\
&=&
- e^2 \sum_{n \kvec} f_n \partial_b \R[ \bra{\partial_a u_{n\kvec} } Q_n \ket{\partial_i u_{n\kvec}} ] q_a q_b \nonumber \\
&=&
- e^2 \sum_{n \kvec} f_n \partial_b g^{ai}_n q_a q_b
\end{eqnarray}

Next, in the case of (ii),
\begin{eqnarray}
\Phi^{i,ab(\mathrm{B-ii})}_{Jn}
&=&
e^2 \sum_{m\neq n \kvec} \frac{f_n}{\epsilon_{nm\kvec}} \biggl\{
- \partial_i \tilde{\epsilon}_{nm\kvec} \R [ \braket{u_{n\kvec} | \partial_a u_{m\kvec}} \braket{u_{m\kvec} | \partial_b u_{n\kvec}} ]
+
\partial_b \tilde{\epsilon}_{nm\kvec} \R[ \braket{u_{n\kvec} | \partial_a u_{m\kvec}} \braket{u_{m\kvec} | \partial_i u_{n\kvec}} ]
\nonumber \\
&&
+ \sum_{l(\neq n)} \epsilon_{ml\kvec} \Bigl(
\R [\braket{u_{n\kvec} | \partial_a u_{m\kvec}} \braket{\partial_b u_{m\kvec} | u_{l\kvec}} \braket{u_{l\kvec} | \partial_i u_{n\kvec}} ] 
+
\R [ \braket{u_{n\kvec} | \partial_a u_{m\kvec}} \braket{u_{m\kvec} | \partial_i u_{l\kvec}} \braket{ u_{l\kvec} | \partial_b u_{n\kvec}} ]
\Bigr) \biggr\} q_a q_b \nonumber \\
&=&
e^2 \sum_{m\neq n \kvec} \frac{f_n}{\epsilon_{nm\kvec}} \Bigl\{
\R[ \A^a_{nm} V^i_{ml,n} \A^b_{ln}] - \R [\A^a_{nm} V^b_{ml,n} \A^i_{ln}]
\Bigr\} q_a q_b
\end{eqnarray}
Here, we define $V_{ml,n}^{i} = \frac{1}{2} (v^{ml}_{i} + v^{0n}_{i} \delta_{ml}) $.

Finally, using Eq.~(\ref{Jn_me_response}), we obtain the orbital ME tensor $\chi^{\mathrm{(me)}}_{\gamma \alpha}=-2i\tilde{\beta}_{\gamma \alpha}=\lim_{\omega \to 0} \varepsilon_{\alpha \beta \delta} ( \Phi^{\beta,\gamma \delta}_{Jn}(\omega) +  \Phi^{\beta,\delta \gamma}_{Jn}(\omega))/3$
\begin{eqnarray}
\chi^{(\mathrm{me})}_{\gamma \alpha}
=
e^2\sum_{n\kvec} \Biggl[ \frac{-f'_n}{\delta} 
\Bigl\{ (\partial_{\gamma} \epsilon_n) m^{\alpha}_n
-\frac{1}{3} \delta_{\gamma \alpha} (\partial_{\beta} \epsilon_n ) m^{\beta}_n
\Bigr\}
+ f_n 
\bigg\{
\frac{-1}{3} \varepsilon_{\alpha \beta \delta} \partial_{\delta} g^{\beta \gamma}_n
+
\sum_{m(\neq n)}
\frac{2}{\epsilon_{nm}} \R \Bigl[ 
\A_{nm}^{\gamma} M_{mn}^{\alpha}
-
\frac{1}{3} \delta_{\gamma \alpha} \A_{nm}^{\beta} M_{mn}^{\beta}
\Bigr]
\biggr\}
\Biggr]. \nonumber \\
\end{eqnarray}
We use the following relation for the orbital magnetic moment
\begin{eqnarray}
\bm{m}_n = \frac{1}{2} \I \bra{\bm{\nabla} u_{n\kvec}} \times (\epsilon_{n\kvec} - H_{\kvec}) \ket{ \bm{\nabla} u_{n\kvec}}.
\end{eqnarray}
This response tensor includes both the extrinsic and intrinsic parts and gives the same result as obtained by using the vector potential in the main text.

\section{Analysis of $\tilde{Q}^{\mathrm{(m)}}_{y'z}(\mu)$ in the Dirac Hamiltonian at zero temperature} \label{dirac_hamiltonian}
We use the Dirac Hamiltonian \cite{PhysRevB.98.060402} (and drop the prime  in the following for simplicity) \begin{eqnarray}
H_{\kvec}^{\mathrm{Dirac}} = v' k_x + v_x k_x \sigma_x + v_y k_y \sigma_y
\end{eqnarray}
for the analysis of $\tilde{Q}^{\mathrm{(m)}}_{yz}(\mu)$ defined in  Sec.~\ref{model_calculation}. In the following, we assume $|v'| < |v_x|, |v_y|$. This simplifies the discussion because the band index changes at $E=0$. The eigenenergies are $\epsilon_{\kvec \pm} = v' k_x \pm \sqrt{(v_x k_x)^2 + (v_y k_y)^2} \equiv v' k_x \pm h_{\kvec}$. At zero temperature, $\tilde{Q}^{\mathrm{(m)}}_{yz}(\mu)$ is given as
\begin{eqnarray}
\tilde{Q}^{\mathrm{(m)}}_{yz}(\mu)
=
e^2 \int \frac{d^2k}{(2\pi)^2} \sum_{n = \pm} \mathcal{G}(\epsilon_{n\kvec}) \biggl(
\frac{1}{3} \varepsilon_{kl z} \partial_l g^{yk}_{n\kvec}
-
\sum_{m(\neq n)} \frac{2}{\epsilon_{nm\kvec}} \R \Bigl[ 
\mathcal{A}^y_{nm} M^z_{mn}
\Bigr]
\biggr), \label{mqm_appendix}
\end{eqnarray}
where we use the grand potential density $ \mathcal{G}(\epsilon_{n\kvec}) = (\epsilon_{n\kvec} - \mu)\Theta(\mu - \epsilon_{n\kvec}) $ at zero temperature. In the following, we consider two cases: contributions from a single band for $\mu < 0$, and (ii) contributions from two bands for $\mu > 0$. We need the off-diagonal Berry connections for the calculation of both cases. The off-diagonal Berry connections are given by
\begin{eqnarray}
\mathcal{A}^x_{+-} = \frac{v_x v_y k_y}{2 h_{\kvec}^2}, \hspace{10pt} \mathcal{A}^y_{+-} = -\frac{v_x v_y k_x}{2 h_{\kvec}^2 }.
\end{eqnarray}
Using these off-diagonal Berry connections, we obtain the quantum metric and the second term in Eq.~(\ref{mqm_appendix}) as
\begin{eqnarray}
&& g^{yx}_{\pm} = \R [\A^y_{\pm \mp} \A^x_{\mp \pm}] = -\frac{1}{4 h_{\kvec}^4} (v_x v_y)^2  k_x k_y \\
&& g^{yy}_{\pm} = \R [\A^y_{\pm \mp} \A^y_{\mp \pm}] = \frac{1}{4 h_{\kvec}^4} (v_x v_y k_x)^2  \\
&& \R[\A^y_{\pm \mp} M^z_{\mp \pm}] = \frac{1}{4 h_{\kvec}^4} v' (v_x v_y k_x)^2.
\end{eqnarray}

At first, we consider the case of (i). $\tilde{Q}^{\mathrm{(m)(-)}}_{yz}(\mu)$ is given by
\begin{eqnarray}
\tilde{Q}^{\mathrm{(m)(-)}}_{yz}(\mu)
&=&
-e^2 \int \frac{d^2 k}{(2\pi)^2} \Theta (\mu - \epsilon_{-\kvec}) \biggl\{
\frac{1}{3} \varepsilon_{klz} v^{0-}_l g^{yk}_{-\kvec} + \frac{2(\epsilon_{-\kvec}-\mu)}{\epsilon_{-+\kvec}} \R[\A^y_{-+} M^z_{+-}] \biggr\} \nonumber \\
&=&
-e^2 \int \frac{d^2 k}{(2\pi)^2} \Theta (\mu - \epsilon_{-\kvec}) \biggl\{
\frac{1}{3} \biggl(
\frac{2v_y^2 k_y}{ h_{\kvec}} \frac{(v_x v_y)^2 k_x k_y}{4h_{\kvec}^4}
-
\biggl(v' - \frac{2v_x^2 k_x}{h_{\kvec}} \biggr) \frac{(v_x v_y k_x)^2}{h_{\kvec}^4}
\biggr)
\nonumber \\
&& \hspace{115pt} - \frac{v' (v_xv_yk_x)^2 (v'k_x -h_{\kvec} -\mu )}{4h_{\kvec}^5} \biggr\}.
\end{eqnarray}
We redefine $|v_x|k_x = h \cos \theta$, $|v_y|k_y = h \sin \theta$ and $|h_{\kvec}| = h$ and rewrite $\tilde{Q}^{\mathrm{(m)(-)}}_{yz}(\mu)$ using these variables as
\begin{eqnarray}
\tilde{Q}^{\mathrm{(m)(-)}}_{yz}(\mu)
&=&
-e^2 \int_0^{\infty} \int_0^{2\pi} \frac{h dh d\theta}{(2\pi)^2 |v_x||v_y|} \Theta (\mu- h(v' \cos \theta/|v_x| -1)) \biggl\{
\frac{v'v_y^2 \cos^2 \theta }{12 h^2} \Bigl( -1 + 3\frac{\mu}{h} \Bigr)
\biggr\}.
\end{eqnarray}
Moreover, we redefine $h$ as $h(v' \cos \theta/|v_x| - 1) = z$ and rewrite $\tilde{Q}^{\mathrm{(m)(-)}}_{yz}(\mu)$ as
\begin{eqnarray}
\tilde{Q}^{\mathrm{(m)(-)}}_{yz}(\mu)
&=&
-e^2 \int_{0}^{-\infty} \int_0^{2\pi} \frac{ dz d\theta}{(2\pi)^2 |v_x||v_y|} \Theta (\mu- z) \biggl\{
\frac{v'v_y^2 \cos^2 \theta }{12 z} \Bigl( -1 + 3\frac{\mu}{z}(v' \cos \theta /|v_x| - 1) \Bigr)
\biggr\} \nonumber \\
&=&
\frac{e^2 v' |v_y|}{16 \pi |v_x| } \biggl( -\frac{1}{3} \log \Bigl| \frac{\mu}{-\Lambda} \Bigr| + 1 \biggr). \label{mu_negative}
\end{eqnarray}
Here, we introduce an energy cutoff $\Lambda$.

Next, we discuss the case of (ii). In this case we need to consider the contribution from the $+$-band. Thus, we will calculate $\tilde{Q}^{\mathrm{(m)(+)}}_{yz}(\mu)$. This quantity is 
\begin{eqnarray}
\tilde{Q}^{\mathrm{(m)(+)}}_{yz}(\mu)
&=&
-e^2 \int \frac{d^2 k}{(2\pi)^2} \Theta (\mu - \epsilon_{+\kvec}) \biggl\{
\frac{1}{3} \varepsilon_{klz} v^{0+}_l g^{yk}_{+\kvec} + \frac{2(\epsilon_{+\kvec}-\mu)}{\epsilon_{+-\kvec}} \R[\A^y_{+-} M^z_{-+}] \biggr\} \nonumber \\
&=&
-e^2 \int \frac{d^2 k}{(2\pi)^2} \Theta (\mu - \epsilon_{+\kvec}) \biggl\{
\frac{1}{3} \biggl(
-\frac{2v_y^2 k_y}{ h_{\kvec}} \frac{(v_x v_y)^2 k_x k_y}{4h_{\kvec}^4}
-
\biggl(v' + \frac{2v_x^2 k_x}{h_{\kvec}} \biggr) \frac{(v_x v_y k_x)^2}{h_{\kvec}^4}
\biggr)
\nonumber \\
&& \hspace{115pt} + \frac{v' (v_xv_yk_x)^2 (v'k_x +h_{\kvec} -\mu )}{4h_{\kvec}^5} \biggr\} 
\end{eqnarray}
We redefine $|v_x|k_x = h \cos \theta$, $|v_y|k_y = h \sin \theta$ and $|h_{\kvec}| = h$ and rewrite $\tilde{Q}^{\mathrm{(m)(+)}}_{yz}(\mu)$ using these variables as
\begin{eqnarray}
\tilde{Q}^{\mathrm{(m)(+)}}_{yz}(\mu)
&=&
-e^2 \int_0^{\infty} \int_0^{2\pi} \frac{h dh d\theta}{(2\pi)^2 |v_x||v_y|} \Theta (\mu - h(v' \cos \theta /|v_x| + 1) )
\biggl\{
\frac{v' v_y^2 \cos^2 \theta}{12 h^2} \Bigl(  
-1 - 3 \frac{\mu}{h}
\Bigr)
\biggr\}.
\end{eqnarray}
Moreover, we redefine $h$ as $h(v' \cos \theta/|v_x| + 1) = z$ and rewrite $\tilde{Q}^{\mathrm{(m)(+)}}_{yz}(\mu)$ as
\begin{eqnarray}
\tilde{Q}^{\mathrm{(m)(+)}}_{yz}(\mu)
&=&
-e^2 \int_{+\delta}^{\infty} \int_0^{2\pi} \frac{dz d\theta}{(2\pi)^2 |v_x||v_y|} \Theta (\mu -z)
\biggl\{
\frac{v' v_y^2 \cos^2 \theta}{12z} \Bigl( 
-1 - 3\frac{\mu}{z} (v' \cos \theta /|v_x| + 1)
\Bigr)
\biggr\} \nonumber \\
&=&\frac{e^2v'|v_y|}{16\pi |v_x|} \biggl(
\frac{1}{3} \log \Bigl| \frac{\mu}{\delta}  \Bigr| -1 + \frac{\mu}{\delta}
\biggr).
\end{eqnarray}
Here, we use $\delta = +0$. In addition to the contribution from the $+$-band, the one from the $-$-band is
\begin{eqnarray}
\tilde{Q}^{\mathrm{(m)(-)}}_{yz}(\mu)
&=&
-e^2 \int_{-\delta}^{-\infty} \int_0^{2\pi} \frac{ dz d\theta}{(2\pi)^2 |v_x||v_y|} \biggl\{
\frac{v'v_y^2 \cos^2 \theta }{12 z} \Bigl( -1 + 3\frac{\mu}{z}(v' \cos \theta /|v_x| - 1) \Bigr)
\biggr\} \nonumber \\
&=&
\frac{e^2v' |v_y|}{16\pi |v_x|} \biggl(
\frac{1}{3} \log \Bigl| \frac{-\Lambda}{-\delta} \Bigr| +\frac{\mu}{-\delta}
\biggr).
\end{eqnarray}
Thus, the total contribution in the case of (ii) is
\begin{eqnarray}
\tilde{Q}^{\mathrm{(m)(-)}}_{yz}(\mu) + \tilde{Q}^{\mathrm{(m)(+)}}_{yz}(\mu)
&=&
\frac{e^2v' |v_y|}{16\pi |v_x|} \biggl(
\frac{1}{3} \log \Bigl| \frac{\mu \Lambda}{\delta^2} \Bigr| -1 
\biggr). \label{mu_positive}
\end{eqnarray}

Comparing Eq.~(\ref{mu_negative}) and Eq.~(\ref{mu_positive}), we can see that there is a jump at the Dirac point ($\mu = 0$) with an additional logarithmic dependence. This behavior induces a peak structure in the intrinsic orbital ME effect.

\end{widetext}
\end{document}